\documentclass{aa}  

\usepackage{graphicx}

\usepackage{txfonts}
\newcommand{\gaia}{\textit{Gaia} }
\defcitealias{Drimmel2023}{D23}
\begin{document}

   \title{Detection of the Keplerian decline in the Milky Way rotation curve}


   \author{Yongjun Jiao\inst{1}, François Hammer\inst{1},
Haifeng Wang\inst{2},
Jianling Wang\inst{1,3},
Philippe Amram\inst{4},
Laurent Chemin\inst{5}, \and
Yanbin Yang\inst{1}
          }

   \institute{GEPI, Observatoire de Paris, Paris Sciences et Lettres, CNRS, Place Jules Janssen 92195, Meudon, France\\
   \email{yongjun.jiao@obspm.fr}
   \and
Centro Ricerche Enrico Fermi, Via Panisperna 89a, I-00184 Rome, Italy
\and
CAS Key Laboratory of Optical Astronomy, National Astronomical Observatories, Beijing 100101, China
\and
Aix-Marseille Univ., CNRS, CNES, LAM, 38 rue Frédéric Joliot Curie, 13338 Marseille, France
\and
Instituto de Astrofisica, Universidad Andres Bello, Fernandez Concha 700, Las Condes, Santiago, RM, Chile
}

   \date{Received 20 July 2023; accepted 21 August 2023}

  \abstract{
Our position inside the Galactic disc has previously prevented us from establishing an accurate rotation curve (RC). The advent of \textit{Gaia} and its third data release (\textit{Gaia} DR3) made it possible to specify the RC up to twice the optical radius. We aim to establish a new RC of the Galaxy from the \gaia DR3 by drastically reducing systematic uncertainties. Our goal is to provide a new estimate of the mass of the Galaxy. We compared different estimates, established a robust assessment of the systematic uncertainties, and addressed differences in methodologies, particularly regarding distance estimates. \\
We find a sharply decreasing RC for the Milky Way; the decrease in velocity between 19.5 and 26.5 kpc is approximately 30 km\,s$^{-1}$. We identify, for the first time, a Keplerian decline of the RC, starting at $\sim$ 19 kpc and ending at $\sim$ 26.5 kpc from the Galaxy centre, while a flat RC is rejected with a significance of 3$\sigma$. The total mass is revised downwards to $2.06^{+0.24}_{-0.13}\times 10^{11}\ M_{\sun}$, which is in agreement with the absence of a significant mass increase at radii larger than 19 kpc. We evaluated the upper limit on the total mass by considering the upper values of velocity measurements, which leads to a strict, unsurpassable limit of $5.4\times 10^{11}\ M_{\sun}$.
}


   \authorrunning{Jiao et al.}
   \maketitle
%

\section{Introduction}

Almost a century ago, \citet[see their Figure 4]{Lundmark1925} was amongst the first to identify the flat behaviour of disc-galaxy rotation curves (RCs). From optical spectroscopy, \citet[then \citealt{Mayall1951}]{Babcock1939} reported that the RC of M31 shows no decrease up to 20 kpc from the centre. With a larger sample of galaxies, \cite{Rubin1978} found that several spiral galaxies have a flat RC. The advent of radio astronomy made it possible to probe galaxy rotation beyond the optical disc. The increased sensitivity of radio telescopes allowed \cite{Bosma1978} to obtain the first sample of galaxies observed in the neutral hydrogen line, and he demonstrated that most galaxies show a flat RC. Extended flat RCs of spiral galaxies can be considered as major evidence of the presence of an extended halo of dark matter (DM) surrounding them. 

Early RC investigations were made beyond our own Galaxy because our location in it and the extinction prevented straightforward, direct determination of the RC as is possible for nearby external galaxies. HII regions, OB stars, carbon stars, planetary nebulae, and cepheids have been used to trace the rotation of the Galaxy. As reviewed by \cite{Schmidt1965}, before the early 1960s, the outer RC was thought to be Keplerian, and from the late 1960s until today, evidence has been found to show that its outer curve is rather flat, albeit with significant uncertainties. Therefore, the Galaxy must contain large amounts of dark matter. However, it is interesting to note that the RC collected by \citet{Sofue2009} is, despite huge uncertainties, consistent with a decreasing RC from $\sim$15 to $\sim$23 kpc.

A revolution in this area came with \textit{Gaia}, whose proper motion measurements have allowed 3D velocity measurements. Pioneering efforts to establish the Milky Way (MW) RC with \gaia DR2 were made by \citet{eilers2019}, who used spectrophotometric parallax distances from \citet{Hogg2019} for a set of more than 20 000 red giant branch (RGB) stars. \citet{eilers2019} provided by far the most accurate RC for the MW, a success confirmed by \citet{mroz2019} on the basis of a small sample of variable stars that nevertheless led to very accurate distance estimates.

The MW has been found to have an exceptionally quiet merger history, evidenced by its rather pristine halo and small angular momentum when compared to other spiral galaxies \citep[and references therein]{Hammer2007}. This was confirmed by the discovery of its last major merger, Gaia--Sausage--Enceladus \citep[GSE, 8-10 Gyr ago,][]{Belokurov2018,Haywood2018,Helmi2018}, which was identified from the angular momentum signatures of residual stars in the halo. Because most spiral galaxies underwent their last major merger more recently \citep[$\sim$ 6 Gyr ago,][]{Hammer2009,Puech2012},  the MW disc may be less affected by the large-scale motions expected following such a major event. This would make the MW one of the most appropriate targets for deriving a RC, assuming equilibrium conditions. It has been argued that a minor merger, such as the infall of Sagittarius (Sgr) 4-6 Gyr ago, could have perturbed the MW outer disc, providing a possible explanation for the warp \citep{Bailin2003} and the observed the vertical oscillations \citep{Laporte2018}. Being at first passage, the Large Magellanic Cloud (LMC) may have affected the MW disc location with respect to its halo \citep{Conroy2021,Erkal2021}, but is unlikely to have affected its internal dynamics or morphology \citep{Laporte2018}.

\gaia DR3 \citep{gaiadr3} has provided improved parallaxes and proper motions, whose systematic uncertainties are smaller by a factor of 2 when compared to those of \gaia DR2 \citep{Lindegren2021a,Lindegren2021b}.  \citet[][hereafter \citetalias{Drimmel2023}]{Drimmel2023}  measured a robust RC of the MW out to $R=14$ kpc from the 3D velocity space of a clean sample of OB stars and another sample of RGB stars using \gaia DR3 data. This was followed by three studies investigating the outer MW RC \citep{wang2023,zhou2023,ou2023}, which used different star samples and methodologies. \citet{wang2023} derived distances from the \gaia parallaxes ($\pi$) of a very large number of stars for which there are also radial velocities from \gaia DR3. These latter authors implemented a Lucy's inversion method \citep[LIM,][]{lucy1974}, from which they populated 6D phase space cells ($\rm l$, $\rm b$, $\pi$, $\rm V_{\rm r}$, $\mu_{\alpha}^{*}$, $\mu_{\delta}$). These authors then determined the average values of the velocity components and their dispersions, which were tested and shown to provide reliable results. \cite{ou2023} followed a technique similar to that of \cite{eilers2019}, namely using spectrophotometric distances to establish the RC of over 30 000 RGB stars. A similar approach was adopted by \citet{zhou2023}, although their distances were estimated using different priors.

\begin{figure}
    \centering
    \includegraphics[width=\columnwidth]{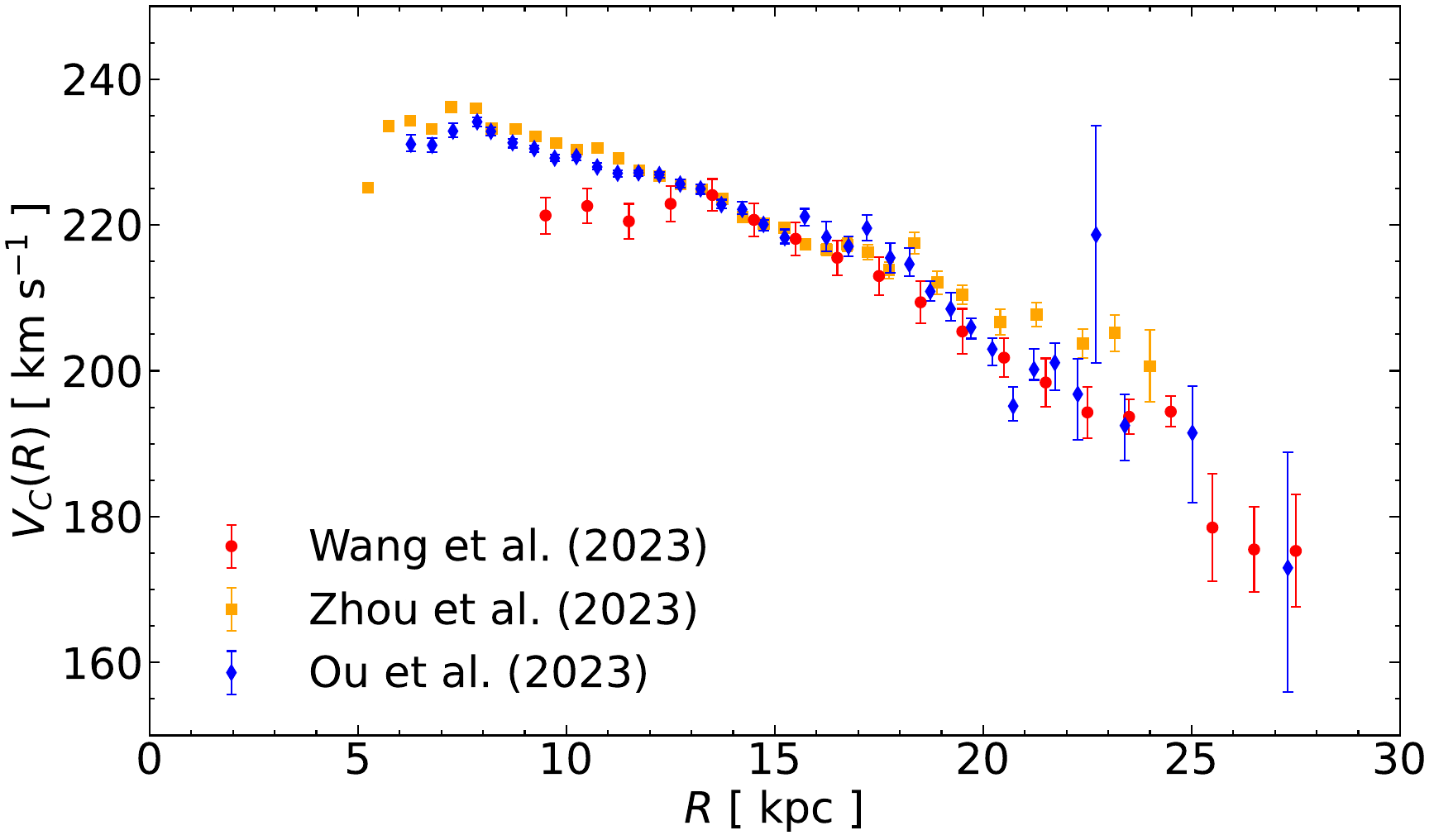}
    \caption{Comparison of the three different measurements of the MW RC based on \gaia DR3.}
    \label{fig:rc_compa}
\end{figure}

Figure~\ref{fig:rc_compa} compares the three corresponding RCs; they show many consistencies, except in the inner (where \citealt{wang2023} have lower velocities) and outer (where \citealt{zhou2023} have higher velocities) parts. This latter discrepancy should significantly affect the estimation of the dynamical mass\footnote{In the following, we use the term dynamical (or total) mass, ($M_\mathrm{dyn}$), to refer to the sum of the baryonic and dark matter masses. This latter is also referred to as the virial dark-matter mass ($M_\mathrm{vir}$) inside the virial radius, above which no additional mass is expected. } of the MW. The goal of the present paper is to verify whether these differences can be attributed to different methodologies in order to establish the MW RC and to determine the dynamical mass of our Galaxy from \gaia DR3. In Section~\ref{sec:method}, we describe the Jeans equation, and show how we derive the systematic uncertainties for the \citet{wang2023} RC using a method similar to that of \citet{eilers2019} and to that of \citet{ou2023}. In Section~\ref{sec:MWmodels}, we show the approach we take to determine the MW dynamical mass using a set of models including baryonic and DM components. In Section~\ref{sec:results}, we show that the three RCs can be reconciled together, providing a sharp velocity drop at large radii, and we provide the MW dynamical mass range and its uncertainty. In Section~\ref{sec:discussion}, we describe the limitations of our estimates, and compare them to estimates based on other tracers. We also test whether or not we can detect the Keplerian decline of the MW RC.

\section{Methods: RC determination and associated uncertainties}
\label{sec:method}

\subsection{The RC derived from Jeans equation}

Assuming an axisymmetric MW potential and a disc at equilibrium, \citet{wang2023} used the Jeans equation \citep[Eq. 4.222a]{Binney2008} to measure the circular velocity curve:

\begin{equation}
    \frac{\partial \nu \langle V_R \rangle}{\partial t} + \frac{\partial \nu \langle V_R^2 \rangle}{\partial R} + \frac{\partial \nu \langle V_RV_z \rangle}{\partial z} + \nu \left( \frac{\langle V_R^2 \rangle -\langle V_\phi^2 \rangle}{R} + \frac{\partial \Phi}{\partial R} \right) = 0,
    \label{eq:jeans}
\end{equation}

where $\nu$ denotes the matter density distribution. By assuming a steady state, a disc that is  symmetric about its equator ($\partial \nu / \partial z = 0$ at $z=0$), and an exponential radial profile of the tracer population with a scale length of $h_R=2.5$ kpc \citep{juric2008}, the circular velocity curve can be derived using $R(\partial \Phi/\partial R)=V_c^2$ in Eq.~\ref{eq:jeans}:

\begin{equation}
    V_c^2=\langle V_\phi^2 \rangle + \frac{R-h_R}{h_R} \langle V_R^2 \rangle -R\frac{\partial \langle V_R^2 \rangle}{\partial R} - R\frac{\partial \langle V_RV_z \rangle}{\partial z}
    \label{eq:rc}
.\end{equation}

For the three velocity components, we apply:
\begin{equation}
    \langle V_X^2 \rangle = \langle V_X \rangle^2 + \sigma_{\langle V_X \rangle}^2
,\end{equation}
where $X$ presents $R$, $\phi,$ and $z$, respectively.

\begin{figure}
    \centering
    \includegraphics[width=\columnwidth]{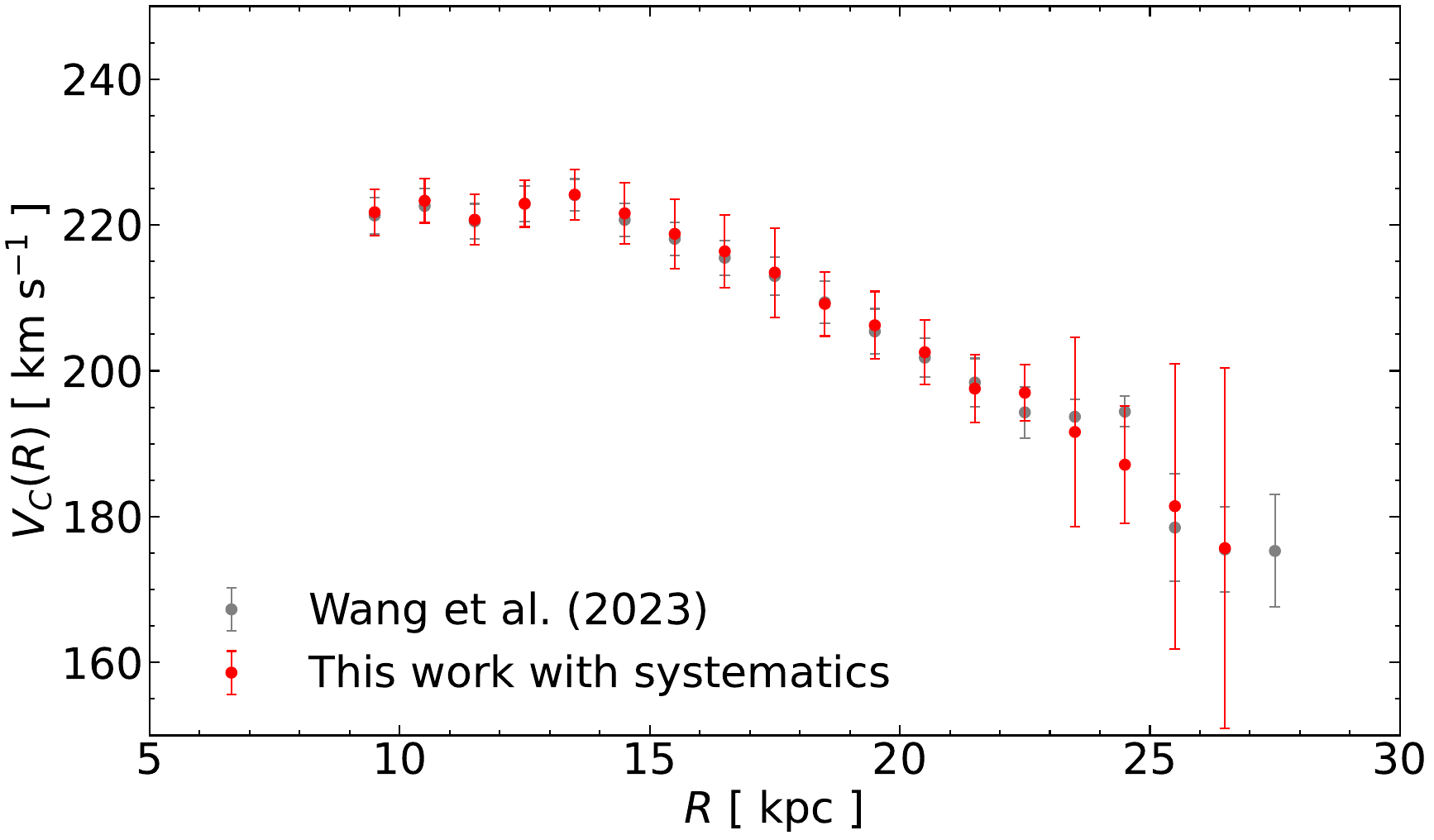}
    \caption{Red points and error bars represent the RC measurement with adjusted bin sizes and systematic uncertainties at a vertical height of $|z| < 3$ kpc. Grey points and error bars present previous findings from \citet{wang2023} without systematic uncertainties.}
    \label{fig:rcs}
\end{figure}

Concerning the RC of \citet{wang2023}, we adjusted the bin size to improve the calculation within each bin. Specifically, we calculated $\langle V_\phi^2 \rangle$ and $\langle V_R^2 \rangle$ using the same 1 kpc width bins as was done by \citet{wang2023}. However, we derived the gradient term $\partial \langle V_R^2 \rangle / \partial R $ after centring the 1 kpc bin width around each data point. Figure~\ref{fig:rcs} shows that the slightly modified RC is consistent with that of \cite{wang2023} within 22 kpc, which means that systematic uncertainties associated with the choice of the bin size and position are very small. The largest modification is for the point at $R=24.5$ kpc, whose velocity amplitude decreases by $8\ \mathrm{km\ s^{-1}}$. However, this velocity is more consistent with the decreasing slope of the RC between 13 and 23 kpc. Data beyond 27.5 kpc are lacking (within a  height of 3 kpc, i.e. $|z|<3$ kpc), and so we do not derive the last point at $R=27.5.$ kpc.

We derived the RC and the associated systematic uncertainties within the height of $3$ kpc, i.e., $|z| < 3$ kpc, which provides the most extended RC in \cite{wang2023}. The cross term $\partial \langle V_RV_z \rangle / \partial z $ was also neglected in the calculations of several previous works \citep[e.g.][]{eilers2019,wang2023} because it is generally considered to be two to three orders of magnitude smaller than the remaining terms at many radii. This term is further analysed in Section \ref{sec:syst_ct}. 

\subsection{Systematic uncertainties}
\label{sec:syst}
The MW has various and complex dynamical structures, which implies that the assumptions of the time-independent gravitational potential and of a smooth density distribution are rough approximations of the true dynamics. For example, the observed velocity field in our Galaxy has been found to have asymmetrical motions with significant gradients in all velocity components \citepalias{Drimmel2023} because of various gravitational influences, such as those of the bar, bulge, and spiral arms, or because of the tidal interaction with the Sagittarius dwarf galaxy \citep{Bailin2003}. It is crucial to estimate the systematic uncertainties brought to the circular velocity curve from the data. \cite{wang2023}  derived the kinematic maps of different velocity components from \gaia DR3 up to 30 kpc. However, the perturbations due to the radial velocity component are sufficiently small, which justifies the use of the time-independent Jeans equation.

\subsubsection{The neglected cross-term}
\label{sec:syst_ct}
\begin{figure}
    \centering
    \includegraphics[width=\columnwidth]{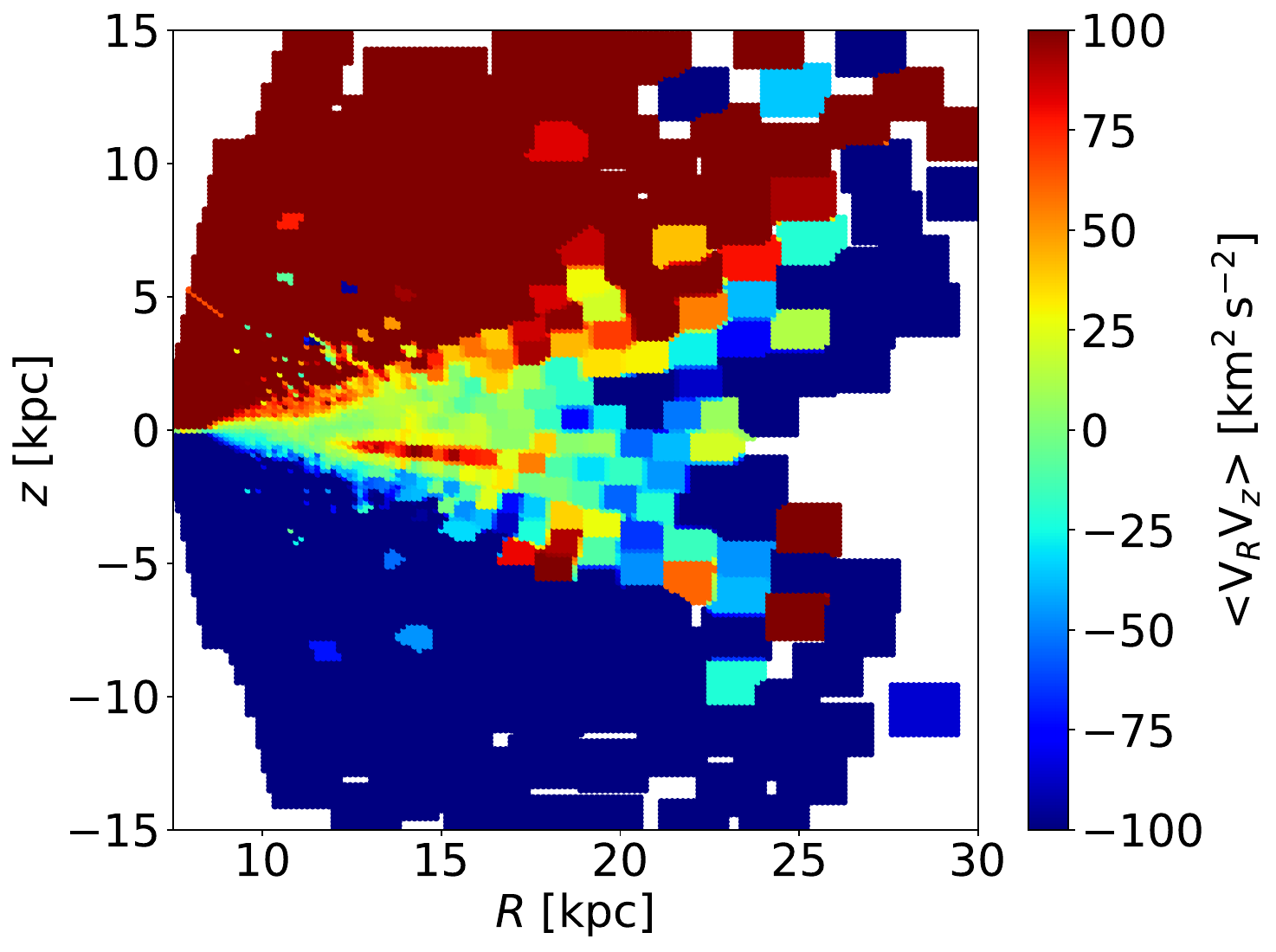}
    \caption{Cross-term $\langle V_RV_z\rangle$ using LIM in the Galactic (R, z) plane.}
    \label{fig:map_ct}
    \end{figure}

The vertical gradient of the cross-term $\langle V_RV_z \rangle$ is usually considered negligible within about 20 kpc. As we consider the RC out to 27 kpc, the largest contribution to the systematic uncertainty of the circular velocity comes from the neglected cross-term in the Jeans equation at large radii. Figure~\ref{fig:map_ct} shows the map of this cross-term on the projection of the Galactic (R, z) plane, which suggests significant variations of the cross-term with radius. The red dotted curve of Fig.~\ref{fig:syst} presents the corresponding contribution to systematic uncertainties. This term causes systematic uncertainties of smaller than $2\%$ below 23 kpc, and of up to $\sim 8\%$ at 25.5 kpc. It should not be neglected at large radii. In addition, we also limited the analysis to a narrower Galactic plane, that is $|z| < 2$ kpc. In this case, systematic uncertainties from this term are smaller than $5\%$ within 25 kpc. 

\subsubsection{Disc scale length}
Another contribution to the systematic uncertainty that cannot be ignored is the unknown density profile of the tracer population, particularly in the outer disc. Our calculation assumes an exponential density profile with a scale length of $h_R=2.5$ kpc. Following \citet{eilers2019}, we vary this scale length, that is, $\Delta h_R = \pm 1$ kpc, and this causes systematic uncertainties at the $\sim1\%$ level, which are represented in Fig. \ref{fig:syst} by an orange-dashed line.

\subsubsection{Disc radial-density profile}
The functional form chosen for the density profile can also lead to systematic uncertainties. We therefore applied a power-law density profile instead, for which we chose an index of $\alpha=-2.25$, which has the same slope as the exponential function at the Sun's location. This leads to a systematic error of less than $4\%$ (blue-dashed line in Figure~\ref{fig:syst}).

\subsubsection{Splitting data sample}
\begin{figure}
    \centering
    \includegraphics[width=\columnwidth]{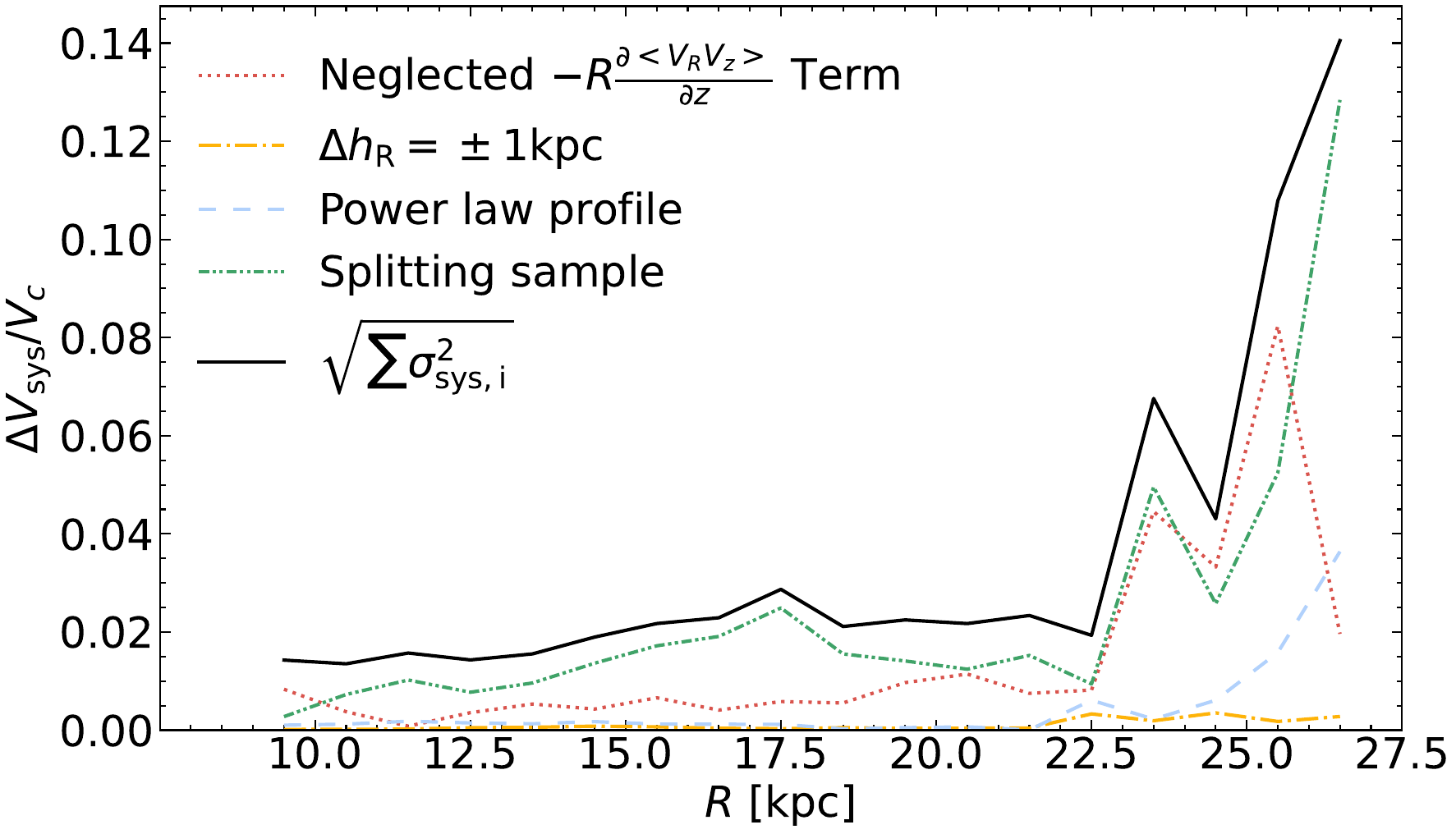}
    \caption{Summary of potential systematic uncertainties in the circular velocity curve at a vertical height of $| z | < 3$ kpc. We estimate systematic uncertainties arising from the neglected cross-term (see the red-dotted line), from varying the exponential scale length of the density profile (orange-dashed line), from passing from an exponential to a power law (blue-dashed line) with an index $\alpha$= -2.25, and finally after splitting the sample into two parts (green-dotted line). The total systematic uncertainty (black-solid line) is at the $2\%$ level up to $R= 22.5$ kpc, and then increases at larger radii.} 
    \label{fig:syst}
\end{figure}

In order to estimate global systematic uncertainties arising from the data sample of different azimuthal ranges, we divided the data sample into two subsamples selected within $160\degr \lid l \lid 180\degr $ and $180\degr \lid l \lid 200\degr $, respectively. We calculated the average of the velocity difference between these two subsamples and the total sample. We present this systematic contribution in Figure~\ref{fig:syst} with a green-dotted line. The systematic uncertainty due to the data sample is smaller than $2\%$ within 22 kpc. At large radii, it is comparable to the systematic uncertainties associated with the neglected cross-term.

\subsubsection{Total uncertainties including systematics}
\label{sec:global_sys}
The total relative systematic uncertainties are indicated by the black-solid line of Figure~\ref{fig:syst}, which is obtained from the quadratic sum of all systematic uncertainties assuming that the errors are all Gaussian:

\begin{equation}
    \sigma_\mathrm{sys}^2=\sigma^2_\mathrm{Cross Term}+\sigma^2_\mathrm{ScaleLength}+\sigma^2_\mathrm{Density Profile} +\sigma^2_\mathrm{SplittingSample}
.\end{equation}

This total has been adopted to calculate error bars in Figure~\ref{fig:rcs}. The latter (see red vertical error bars) corresponds to the quadratic sum of statistical errors with systematic uncertainties. 

However, we note that the systematic uncertainty due to the neglected cross-term and that obtained after splitting the sample into two parts follow the same trend (compare the red-dotted with the green-dashed line in Figure~\ref{fig:syst}). A quite similar trend may be identified in Fig. 5 of \citet{ou2023}. This leads us to suspect that systematic uncertainties are overestimated by both studies.

\section{Methods: MW mass models}
\label{sec:MWmodels}

\subsection{Varying baryonic models for bulge and disc}
\label{sec:bary}

The contribution of the baryonic components to the MW mass is still uncertain, and this may affect the determination of the DM distribution \citep{karukes2020,jiao2021}. The basic idea is to cope with uncertainties in baryons by using a very large grid of possible models. However, we are aware that some baryonic models may not be fully consistent with other important constraints from the vertical dynamics of the disc stars \citep{bovy2013} or from microlensing \citep{wegg2016}.

To verify how the baryonic mass distribution may affect the RC, we followed \cite{jiao2021} in considering several possible combinations of models for the bulge and the disc presented by \cite{iocco2015}, all with baryonic mass smaller than $7 \times 10^{11} M_{\sun}$\footnote{We notice that Model I of \citet{Pouliasis2017} used by \citet{eilers2019} and \citet{jiao2021} overestimates the bulge mass (Paola di Matteo, 2022, private communication).}. In addition, we also considered the baryonic model B2 from \cite{desalas2019}, which includes neutral gas and dust components, and was used by \citet{ou2023}. Two triaxial density profiles E (Exponential-type) and G (Gaussian-type) for the bulge from \cite{stanek1997} can be expressed as:

\begin{equation}
\begin{aligned}   
    &\mathrm{E} : \rho_{\mathrm{bulge}}(x,y,z)=\rho_0\ e^{-r}  \\ &\mathrm{G} : \rho_{\mathrm{bulge}}(x,y,z)=\rho_0\ e^{-r^2/2} 
\end{aligned}
\end{equation}
with :
\begin{equation}
    r^2=\frac{x^2}{x_b^2}+\frac{y^2}{y_b^2}+\frac{z^2}{z_b^2}, 
\end{equation}
where $(x,y,z)$ are the coordinates along the major, intermediate, and minor axes. The bulge of B2 is modelled with a Hernquist potential:  
\begin{equation}
    \Phi(r)=-\frac{GM}{r+r_b}
    \label{eq:hernq}
.\end{equation}

For the disc component, we adopted a double exponential as described below:

\begin{equation}
    \rho(R,z)=\rho_0\exp{\left( -\frac{R}{L}-\frac{z}{H} \right)}
\label{eq:doubl_exp}
,\end{equation}
where $\rho_0=M/(4\pi H L^2)$ is the normalisation, $M$ is the corresponding disc mass, and $L$ and $H$ are the disc scale length and height, respectively. Three-disc models (CM from \cite{calchi2011}, dJ from \cite{dejong2010} and J from \cite{juric2008}) contain thin and thick discs. Model  B2 possesses only a thin disc. The dust component and the gas distribution of B2 are also modelled as double exponential profiles (Eq.~\ref{eq:doubl_exp}).

\begin{table}
\centering
\caption{Parameters of bulge, see more details in Section~\ref{sec:bary}}
\begin{tabular}{l l l l}
    \hline \hline
    Parameter & bulge E & bulge G & bulge B2 \\
    \hline
    $M_{\mathbf{bulge}} (10^{10} M_{\odot})$ & 1.962 & 1.639  & 1.550 \\
    $x_\mathrm{b}(\mathrm{kpc})$    &0.899&1.239& 0.700$^a$\\
    $y_\mathrm{b}(\mathrm{kpc})$    &0.386&0.609& ----\\
    $z_\mathrm{b}(\mathrm{kpc})$    &0.250&0.438& ----\\
    \hline
    \multicolumn{4}{l}{$^a$ Value of $r_b$ in Eq.~\ref{eq:hernq}}
    \end{tabular}
\label{tab:bulge}
\end{table}

\begin{table}
\centering
\caption{Parameters of disk, see more details in Section~\ref{sec:bary} }
\begin{tabular}{l l l l l}
    \hline \hline
    Parameter & disk CM  & disk J & disk dJ & disk B2\\
    \hline
    $M_{\mathbf{thin}} (10^{10} M_{\odot})$ & 3.11&3.17&3.33 & 3.65\\
    $M_{\mathbf{thick}} (10^{10} M_{\odot})$ &0.82&0.90&0.78&----\\
    $L_{\mathbf{thin}} (\mathrm{kpc})$ &2.75&2.60&2.60&2.35\\
    $L_{\mathbf{thick}} (\mathrm{kpc})$ &4.10&3.60&4.10&----\\
    $H_{\mathbf{thin}} (\mathrm{kpc})$ &0.25&0.30&0.25&0.14\\
    $H_{\mathbf{thick}} (\mathrm{kpc})$ &0.75&0.90&0.75&----\\
    \hline 
\end{tabular}
\label{tab:disk}
\end{table}

Tables~\ref{tab:bulge} and~\ref{tab:disk} provide the various parameters adopted to represent the bulge and the disc of the MW. 

\subsection{Dark matter model for the halo}

The Einasto profile \citep{einasto1965, retana2012} is widely used to describe the distributions of stellar light and of the mass of dark matter in galaxies, whose density is defined as:
\begin{equation}
    \rho\left(r\right) = \rho_0\  \mathrm{exp}\left[-\left(\frac{r}{h}\right)^{1 / n}\right]
    \label{eq:einasto}
,\end{equation}
where $n$ is the Einasto index, which can determine how fast the density decreases with $r$. Several studies \citep[e.g.][]{chemin2011,jiao2021} have shown that the Einasto profile gives a significantly better fit to the RCs when compared to the Navarro, Frenk \& White profile \citep[NFW,][]{navarro1997}. In particular, \cite{jiao2021} found that the three-parameter Einasto profile may account for a larger range of outer slopes and generate a plausible wide mass range. This is also supported by \citet{ou2023} and \citet{Sylos_Labini2023}, who found the NFW profile unsuitable for reproducing the external slope of the MW RC. Therefore, the present study focuses on a spherical Einasto profile for modelling the MW RC, and Figure~\ref{fig:rc_fit} presents one of our best fits with n=0.43, h=11.41 kpc, and $\rho_{0}=0.01992~M_{\sun}\,\mathrm{pc}^{-3}$ (see first line of Table~\ref{tab:mass}).\\
Constraints on the  shape of the dark matter halo in the Milky Way are weak and no consistent picture has yet emerged. However, at 20 kpc scales, \citet{Kupper2015} and \citet{ Koposov2010} determined the flattening of the dark halo to be $q_z$ = 0.95 $\pm$ 0.15, that is, almost spherical. In the inner regions of the MW, the disc and bulge dominate the RC in that they contribute $\sim85\%$ of the rotational velocity and $\sim70\%$ of the rotational support at 2.2 disc scale lengths, i.e., where the disc velocities are maximal and therefore the disc is maximal \citep{Sackett1997}. The wide variety of baryonic models used in this study allows us to test the hypothesis of the maximum disc and to resolve the debate over the core or cusp nature of the dark halo of our galaxy. It also allows us to test the NFW halo, which in any case cannot fit a decreasing RC.\\
The optical diameter of an external galaxy is often defined using the $D_{\rm 25}$ isophote. This limit has been used to estimate the radius of the MW to 13.4 kpc (e.g. \citealt{Goodwin1998}).  In the following, we obtain a radius that is almost twice as large.
 

\section{Results}
\label{sec:results}
\subsection{Measurement of the rotation curve and comparison with \citet{ou2023}}

\begin{figure}
    \centering
    \includegraphics[width=\columnwidth]{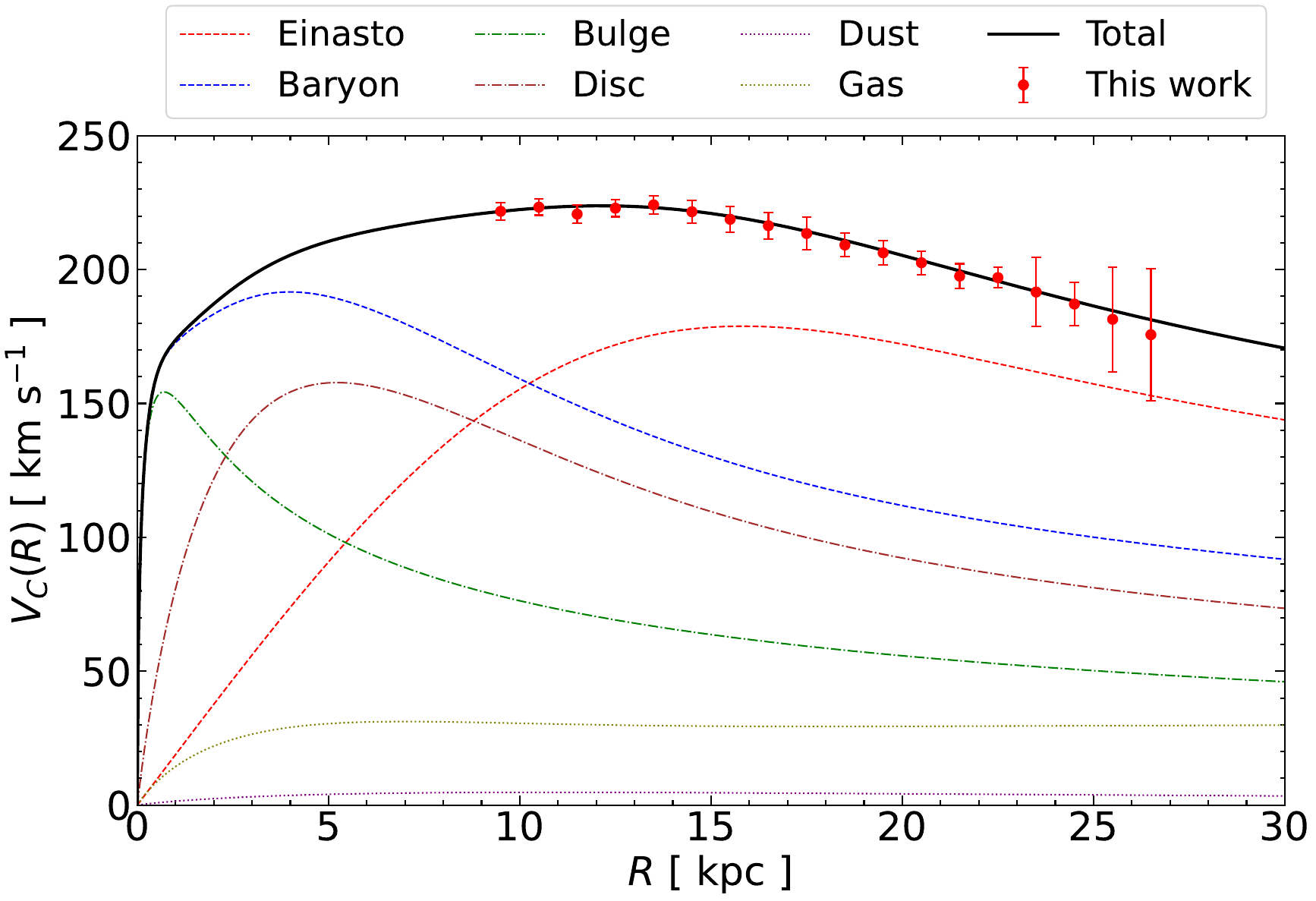}
    \caption{Circular velocity of the Milky Way. The red data points are the measurements computed in this work; error bars include systematic uncertainties. The black solid line represents the sum of the baryonic and dark matter components: the baryonic model B2 (blue-dashed line), including its decomposition into baryonic components (bulge, disc, gas, and dust) and the best fit of  the Einasto dark matter profile (red-dashed line).}
    \label{fig:rc_fit}
\end{figure}


Figure~\ref{fig:rc_fit} shows our final circular velocity curve. In our study, as well as in \cite{ou2023}, error bars account for systematic uncertainties. For R $\ge$ 13 kpc, the two RCs are in reasonably good agreement except for one point at about $23$ kpc (see Figure~\ref{fig:rc_compa}). We suspect that this discrepancy is caused by the disagreement over the radial velocity component at $R\sim 23$ kpc, for which the top panel of Fig. 3 of \cite{ou2023}  shows a large deviation on $\sqrt{\langle v^2_R\rangle}$. In the range of $R=9\sim 13$ kpc, our RC points are slightly lower than those of \cite{ou2023}, which is discussed in Appendix~\ref{AppendixA}.\\
The largest discrepancy between the RC of this paper and that of \cite{ou2023} is perhaps related to the amplitude of the error bars, which are larger in this latter study (compare Figure~\ref{fig:syst} with Figure 5 of \citealt{ou2023}).


\begin{table}
\centering
\caption{Measurements of the Circular Velocity of the Milky Way.}
\begin{tabular}{ccc}
\hline \hline
$R$ [kpc] & $V_c$ [km s$^{-1}$] & $\sigma_{V_c}$ [km s$^{-1}$]\\
\hline
9.5 & 221.75 & 3.17 \\
10.5 & 223.32 & 3.02 \\
11.5 & 220.72 & 3.47 \\
12.5 & 222.92 & 3.19 \\
13.5 & 224.16 & 3.48 \\
14.5 & 221.60 & 4.20 \\
15.5 & 218.79 & 4.75 \\
16.5 & 216.38 & 4.96 \\
17.5 & 213.48 & 6.13 \\
18.5 & 209.17 & 4.42 \\
19.5 & 206.25 & 4.63 \\
20.5 & 202.54 & 4.40 \\
21.5 & 197.56 & 4.62 \\
22.5 & 197.00 & 3.81 \\
23.5 & 191.62 & 12.95 \\
24.5 & 187.12 & 8.06 \\
25.5 & 181.44 & 19.58 \\
26.5 & 175.68 & 24.68 \\
\hline
\end{tabular}
\label{tab:rc_z3}
\end{table}


Both RCs show a significant decline with increasing radius, which can be well approximated by a linear function (see Figure~\ref{fig:rcs}):

\begin{equation}
    V_c\left(R\right) = V \left( R_{\sun} \right) + \beta \left( R-R_{\sun} \right)
    \label{eq:lin_rc}
,\end{equation}
where $R_{\sun}$ is the distance between the Sun and the Galactic centre\footnote{We note that $R_{\sun}=8.34$ kpc for this RC and 8.178 kpc for RC of \citet{ou2023}.}. We find that the slope of our declining RC is $\beta=-(2.18\pm 0.23)\rm\,km\,s^{-1}\,kpc^{-1}$, which is similar to the value of $\beta=-(2.22\pm 0.20)\rm\,km\,s^{-1}\,kpc^{-1}$ obtained by \citet{ou2023}\footnote{In the present study, we accounted for the systematic uncertainties of \citet[see their Fig. 5]{ou2023} when deriving parameters from the corresponding RC. The values that cannot be seen in their Fig. 5 have been chosen to be 0.14.}. 

\cite{wang2023} also split the Galactic region into two, one with galactic latitude b $>0\degr$ and the other with b$<0\degr$ (or one with $z>0$ kpc and the other with $z<0$ kpc) and found an uncertainty on the slope of RC of $\sim 20\%$.

\subsection{Comparison with \citet{zhou2023}}

Figure~\ref{fig:rc_compa} shows that the RC from \citet{zhou2023} indicates larger velocities at the MW disc outskirts. In Appendix~\ref{AppendixB}, we compare the distances adopted by the present study to those adopted in other estimates (see Figure~\ref{fig:dist_compa}), which leads us to suspect that the distances by \citet{zhou2023} are overestimated. After correcting for this, it appears that the \citet{zhou2023} RC is consistent with both the RC of the present study and that from \citet{ou2023}. We also notice that \citet{zhou2023} did not consider the impact of the cross-term when analysing systematic uncertainties. For consistency, we have not considered this study in the following.

\subsection{Estimated range for the dynamical mass of the Milky Way}

Using a Bayesian analysis, one can determine the posterior distribution of the model parameters based on the given data. In the present study, we applied the Markov Chain Monte Carlo (MCMC) affine invariant sampler \textsc{\footnotesize EMCEE} \footnote{\url{https://github.com/dfm/emcee}} \citep{foreman2013}  to test the parameter space of the Einasto profile using flat priors; that is, $M_0=4\pi h^3\rho_0$, $h$, and $1/n$, from $10^{10}$ to $10^{14} M_{\sun}$, from 0 to 20, and from 0 to 5, respectively. Following previous studies, the sum of the logarithm of the likelihood for the observed RC can be derived as:
\begin{equation}
    \ln{\mathcal{L}} = -\frac{1}{2} \sum_i \left( \frac{v_{\mathrm{mod},i}-v_{\mathrm{obs},i}}{\sigma_{i}} \right)^2
    \label{eq:likelihood}
,\end{equation}
where the summation $i$ is done over all the data points, $v_{\mathrm{mod}}$ is the theoretical circular velocity from the MW models, $v_{\mathrm{obs}}$ is the measured circular velocity, and $\sigma$ is the statistical uncertainty of the measurement (see Sect. \ref{sec:syst}).

In the present study, we extrapolated the DM halo to the virial radius, denoted as $R_\mathrm{vir}$, which encloses the virial or the DM contribution to the dynamical mass, $M_\mathrm{vir}$. The virial radius is defined as the radius of a sphere within which the average density of dark matter is equal to 200 times the critical density of the Universe $\rho_\mathrm{cr}$. Here, we adopted a critical density of $\rho_\mathrm{cr}=1.34 \times 10^{-7} M_{\odot}/\mathrm{pc}^3$ \citep{hinshaw2013}.

\begin{figure*}
\begin{minipage}{\columnwidth}
    \centering
    \includegraphics[width=0.95\columnwidth]{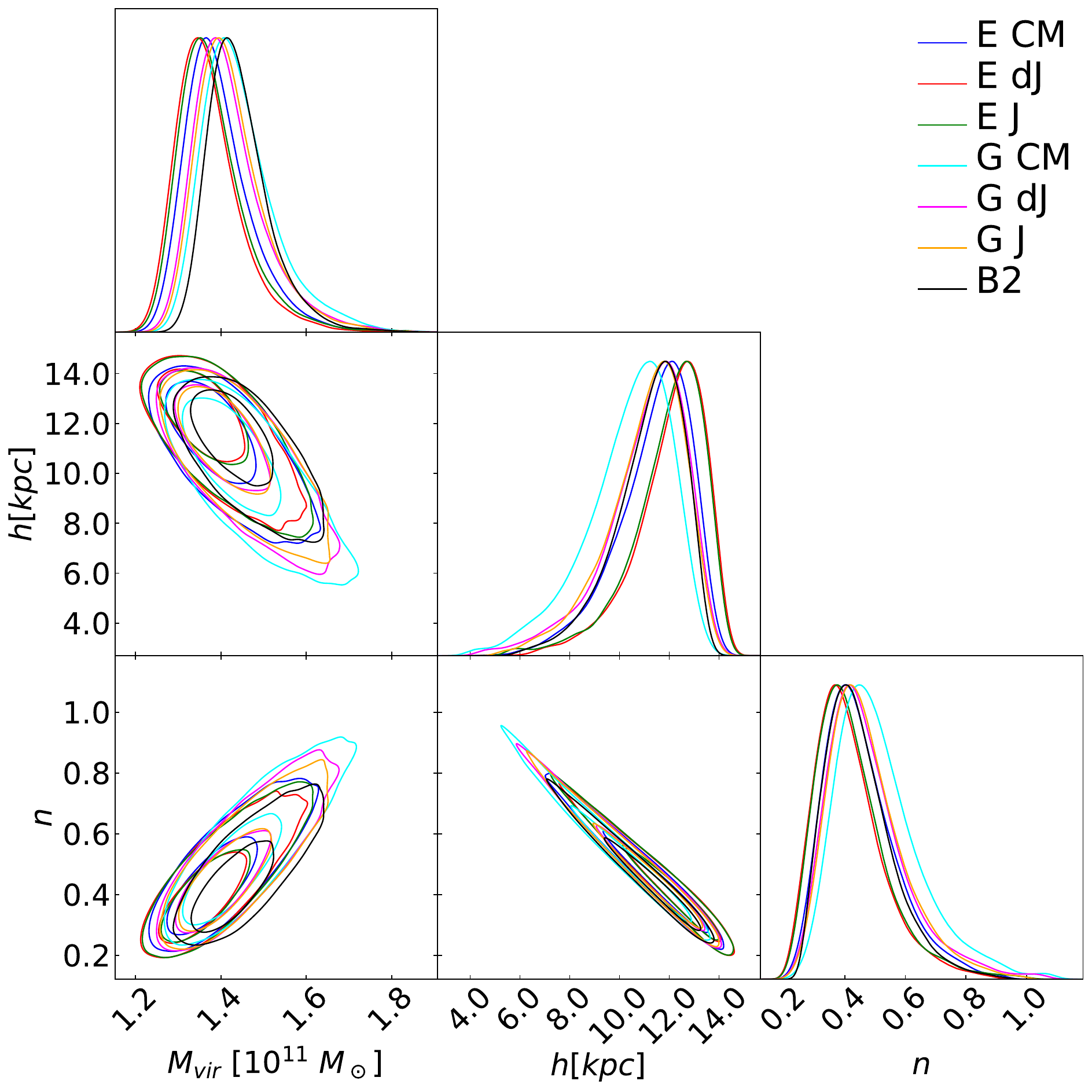}
\end{minipage}
\begin{minipage}{\columnwidth}
    \centering
    \includegraphics[width=0.95\columnwidth]{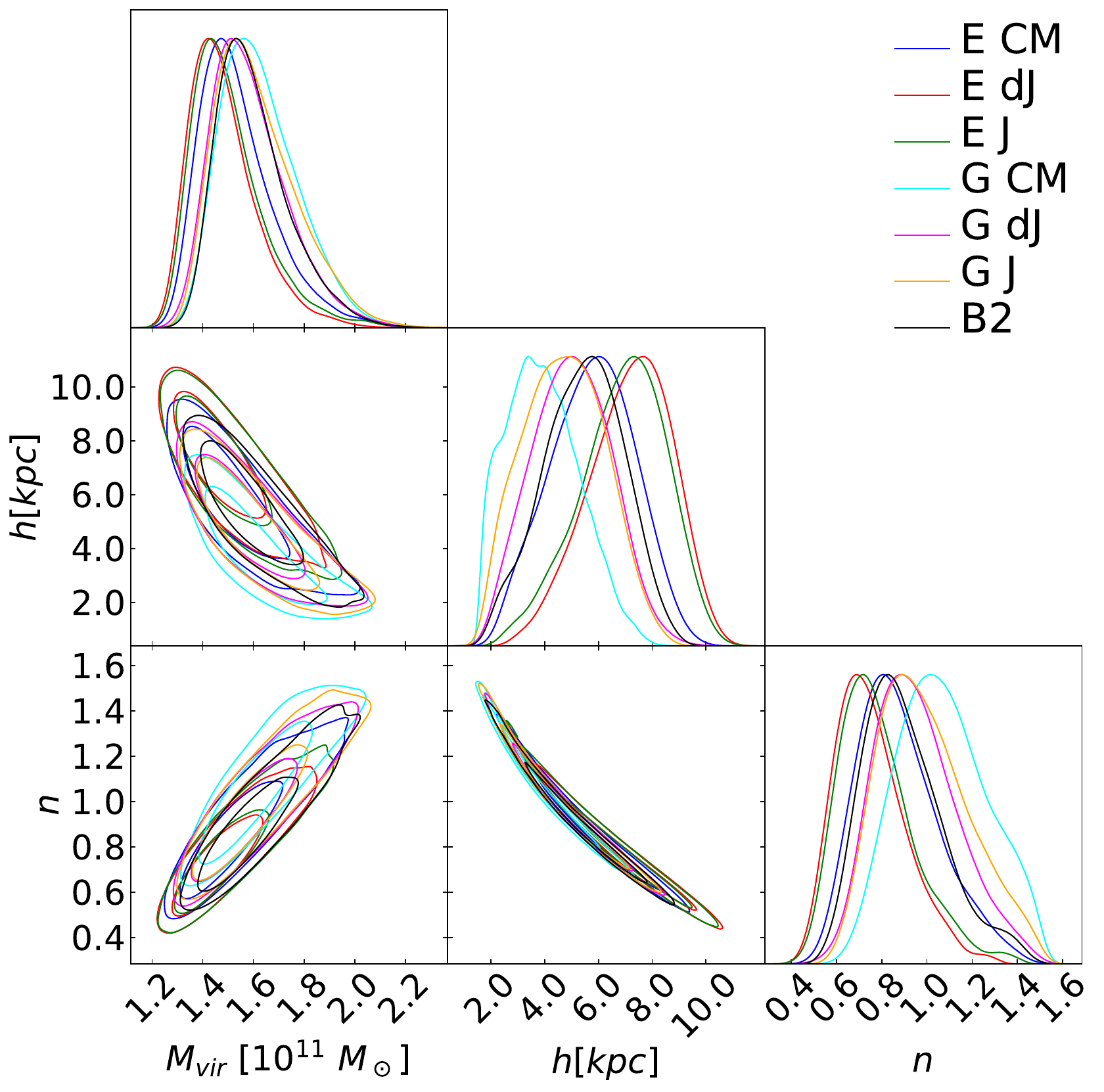}
\end{minipage}
\caption{MCMC tests of Einasto dark matter parameters with several baryonic models (represented by coloured lines) for our RC (right panel) and that of \citet{ou2023} RC.}
\label{fig:mc_dis_eina}
\end{figure*}

In order to properly estimate the dynamical mass, we investigated various baryonic models, as described in Sect. \ref{sec:bary}. The posterior distributions for both RC fits with different baryonic models are given in Fig.~\ref{fig:mc_dis_eina}. We note that we converted the parameters of the virial DM mass $M_\mathrm{vir}$, scale length $h$, and Einasto index $n$.

We calculated the dynamical mass as the sum of the baryonic mass and the DM halo virial mass. Results are presented in Table~\ref{tab:mass}. We find a MW dynamical mass of $1.99^{+0.09}_{-0.06}\times 10^{11}\ M_{\sun}$ at $R_\mathrm{vir}=121.03^{+1.80}_{-1.23}$ kpc from the present work, which can be compared to the dynamical mass of $2.13^{+0.17}_{-0.12}\times 10^{11}\ M_{\sun}$ at $123.80^{+3.21}_{-2.37}$ kpc using the RC of \cite{ou2023}. The local DM density is found in the range of $0.011$ to $0.012~M_{\sun}\,\mathrm{pc}^{-3}$ ($0.418$-$0.456~\mathrm{GeV\,cm}^{-3}$) for both RCs with different baryonic models. One may question the significance of such small error bars, given the fact that they lead to only very small values for the MW dynamical mass. On one hand, this is assuming that each point (and associated error bars) of the RC has been determined independently from the others, which is the underlying assumption of the MCMC method. This is likely true for the RC points, but not necessarily for the error bars, which may have led to an underestimation of the  error bars on dynamical mass. On the other hand, error bars account for systematic uncertainties, and the latter may have been overestimated, as shown in Section~\ref{sec:global_sys}.\\
We performed another test to evaluate the error bars of the MW dynamical mass. To this end, we arbitrarily replaced all the points of our RC by values provided by their upper error bars. By fitting these points with different baryonic models, we obtain dynamical masses ranging from 2.44 to 2.53 $\times\ 10^{11} M_{\sun}$. The latter value would correspond to an absolute upper limit for the MW dynamical mass. However, Figure~\ref{fig:compa_D23} shows that \citet{ou2023} RC points have larger error bars than in our study. Performing the same exercise with these values, one would find total mass values from 3.8 to 5.4 $\times\ 10^{11} M_{\sun}$. We consider the latter value as an absolute upper limit on the MW dynamical mass.

\begin{table*}[h]
    \centering
    \caption{Estimated MW dynamical mass and associated Einasto profile parameters for the RC of the present work (TW) and for that of \citet[][O23]{ou2023}. }
    \begin{tabular}{lccccccccc}
    \hline \hline
    Baryon model & $M_\mathrm{bar}$ & \multicolumn{2}{c}{$M_\mathrm{dyn}$} & \multicolumn{2}{c}{$M_\mathrm{0}$} & \multicolumn{2}{c}{h} & \multicolumn{2}{c}{n} \\
    & [$10^{11} M_{\sun}]$ & \multicolumn{2}{c}{[$10^{11} M_{\sun}$]} & \multicolumn{2}{c}{[$10^{11} M_{\sun}$]} & \multicolumn{2}{c}{[kpc]} & \multicolumn{2}{c}{} \\
    &&TW&O23&TW&O23&TW&O23&TW&O23\\
    \hline
    B2 & 0.616 & $2.05^{+0.08}_{-0.06}$ & $2.19^{+0.17}_{-0.12}$ & $3.72^{+0.45}_{-0.70}$  & $1.23^{+0.63}_{-0.58}$ & $11.41^{+1.15}_{-1.62}$ & $5.5^{+1.46}_{-1.56}$ & $0.43^{+0.12}_{-0.09}$ & $0.87^{+0.20}_{-0.15}$\\
    E dJ & 0.607 & $1.97^{+0.09}_{-0.06}$ & $2.07^{+0.15}_{-0.11}$ & $3.72^{+0.36}_{-0.63}$  & $1.82^{+0.64}_{-0.72}$ & $12.30^{+1.10}_{-1.63}$ & $7.3^{+1.46}_{-1.74}$ & $0.40^{+0.13}_{-0.09}$ & $0.73^{+0.18}_{-0.14}$ \\
    E J & 0.603 & $1.97^{+0.09}_{-0.06}$ & $2.08^{+0.16}_{-0.11}$ & $3.72^{+0.36}_{-0.70}$  & $1.70^{+0.65}_{-0.72}$ & $12.21^{+1.12}_{-1.68}$ & $7.03^{+1.49}_{-1.79}$ & $0.40^{+0.13}_{-0.09}$ & $0.76^{+0.19}_{-0.14}$\\
    E CM & 0.589 & $1.97^{+0.09}_{-0.06}$ & $2.10^{+0.17}_{-0.12}$ & $3.55^{+0.43}_{-0.73}$  & $1.26^{+0.69}_{-0.61}$ & $11.63^{+1.20}_{-1.77}$ & $5.81^{+1.58}_{-1.71}$ & $0.43^{+0.14}_{-0.10}$ & $0.85^{+0.21}_{-0.16}$ \\
    G dJ & 0.575 & $1.98^{+0.09}_{-0.07}$ & $2.14^{+0.17}_{-0.12}$ & $3.47^{+0.51}_{-0.78}$  & $1.02^{+0.64}_{-0.52}$ & $11.34^{+1.28}_{-1.86}$ & $4.99^{+1.52}_{-1.54}$ & $0.45^{+0.15}_{-0.10}$ & $0.94^{+0.21}_{-0.17}$ \\
    G J  & 0.571 & $1.98^{+0.09}_{-0.06}$ & $2.15^{+0.19}_{-0.13}$ & $3.39^{+0.50}_{-0.76}$  & $0.89^{+0.62}_{-0.51}$ & $11.29^{+1.26}_{-1.82}$ & $4.72^{+1.60}_{-1.64}$ & $0.45^{+0.14}_{-0.10}$ & $0.97^{+0.24}_{-0.18}$ \\
    G CM & 0.557 & $1.99^{+0.10}_{-0.07}$ & $2.16^{+0.18}_{-0.13}$ & $3.31^{+0.58}_{-0.91}$  & $0.64^{+0.56}_{-0.38}$ & $10.63^{+1.40}_{-2.01}$ & $3.85^{+1.52}_{-1.35}$ & $0.49^{+0.16}_{-0.11}$ & $1.06^{+0.23}_{-0.19}$ \\
     \hline
     Average & & $1.99^{+0.09}_{-0.06}$ & $2.13^{+0.17}_{-0.12}$\\
    \hline
    \end{tabular}
    \label{tab:mass}
\end{table*}

\section{Discussion}
\label{sec:discussion}

\subsection{The impact of asymmetric drift on the rotation curve measurement}

\begin{figure}
 \centering
 \includegraphics[width=1\columnwidth]{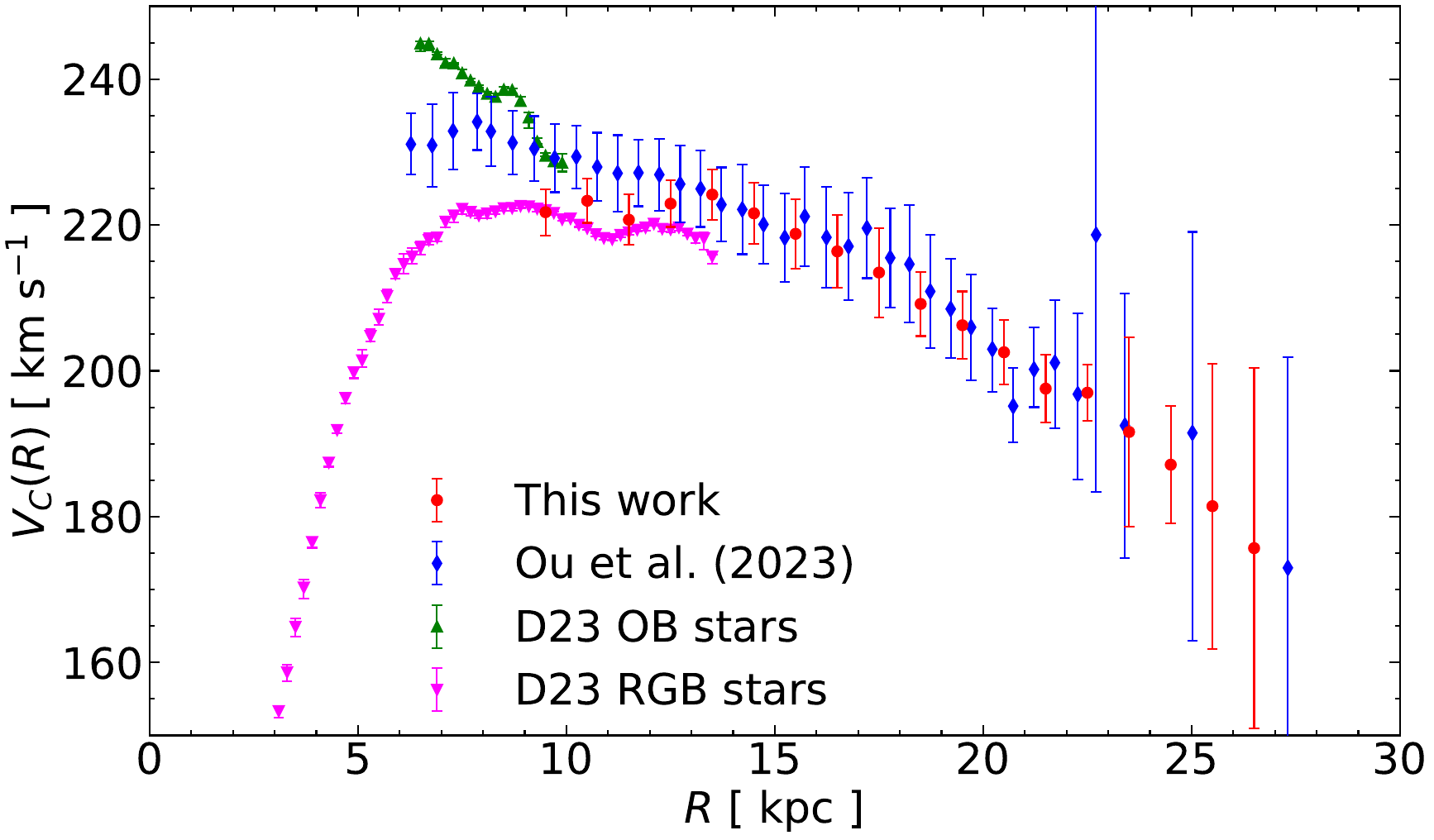}
    \caption{Circular velocity of the Milky Way. The red (blue) data points are the measurements computed in the present work (and in \citealt{ou2023}), with error bars including systematic uncertainties. These are compared with data points from \citetalias{Drimmel2023} (OB: magenta triangles, and RGB: green triangles, stars) that have not been corrected for asymmetric drift.}
\label{fig:compa_D23}
\end{figure}

We calculated the asymmetric drift following Eq. 4.225, Eq. 4.227, and Eq. 4.228 (Str$\ddot{o}$mberg's equation) in \citet{Binney2008},  and it could impact RCs. \citetalias{Drimmel2023} provided velocities for RGB and OB stars that were not corrected for the asymmetric drift. Figure~\ref{fig:compa_D23} compares the corresponding RCs to ours and that of \citet{ou2023}. This comparison shows that the non-corrected RGBs (magenta triangles) from \citetalias{Drimmel2023} lag the corrected curves (see blue and red dots), which clearly illustrates the necessity of the asymmetric drift correction. OB stars (green triangles) are rotating faster, and as they rotate closer to the circular velocity, this would imply that both the correction in our study and that in \citet{ou2023} are sufficient, at least for radii of greater than 10 kpc.\\
However, our correction does not seem entirely sufficient for $R < 10$ kpc, perhaps because of the unknown shape of the true density profile of the MW. This might also mean that the decrease in this region is even steeper than shown by the red and blue points. In the future, extending the very young star rotation curve towards inner disc regions, as well as to larger radii, may provide evidence supporting the decrease in the MW RC, and in particular might help to better determine the radius at which the decrease occurs. 

Concerning systematic uncertainties introduced by the asymmetric drift, \cite[see their Fig. 5]{ou2023} showed that beyond 22 kpc, the asymmetric drift can contribute to relative systematic uncertainties at the level of $15\%$, which might cause the peculiar point around 23 kpc. Figure 8 of \cite{wang2023} shows that the declining slope is shared for stars selected at different heights, for which differences are most likely caused by the asymmetric drift effect. Points near and above $R=24$ kpc might also be affected by the asymmetric drift systematic errors, which is similar to the findings of  \cite{ou2023} for their RC.



\subsection{Does the Milky Way warp have a significant impact on the rotation curve measurement?}

The MW has a warped disc. Observations show that the MW disc is flat out to roughly the Solar Circle, where it then bends upwards and downwards in the northern and southern hemispheres. The amplitude of the warp clearly increases with radius and varies with azimuthal angle. The position and kinematics of the Galactic warp also depend on the stellar populations \citep{poggio2018, wang2020, chrobakov2020, li2023}. 

\cite{wang2020} analysed the stellar warp of different stellar populations by combining LAMOST DR4 and \gaia DR2. These authors concluded that the Galactic warp might not be caused by the gravitational interaction scenarios but by the gas infall process, and gravitational interactions such as that due to the Sgr passages are also addressed.

On the other hand, there has been no major merger in the MW for between 9 and 10 Gyr, which corresponds to the GSE event, as described in Section 1. Interactions with nearby dwarfs or satellites are not expected to have a great impact on disc stability, though Sgr passages near the disc edge have been suggested to have affected or even possibly formed the warp \citep{Bailin2003}. Even in the latter case, if the Sgr main interaction with the disc occurred more than 5 Gyr ago, it is likely that the MW disc is in relative equilibrium (quasi-equilibrium within 30 kpc)\footnote{At 30 kpc, a star rotating at $170~\mathrm{km\,s}^{-1}$ would have the time to make 4.5 orbits, i.e. a sufficiently large number to consider the system to be at equilibrium \citep{Gnedin1999}.} with some oscillations. Those oscillating asymmetries on either side of the galactic plane are called corrugations. From a sample of 40 nearby low-inclination disc galaxies, \citet{Urrejola-Mora2022} identified that $20\%$ of the galaxies exhibit vertically perturbed galactic discs, which could be described by corrugations. One of the most famous corrugations in the MW is the Monoceros Ring located at low Galactic latitudes. Although this is more likely to be explained by disc flaring \citep{Wang2018,Bergemann2018}, scenarios of a perturbed disc by an ancient disrupted dwarf galaxy have also been proposed \citep{Conn2007, Conn2012,Johnston2017}. Corrugations start from the Sun and at least four ripples are seen in the disc outskirts \citep{Newberg2002,Ibata2003,Xu2015}; these blur our representation of a flat disc and can produce wiggles in the RC that are different from those due to spiral arms. 

\cite{chrob2020} used N-body simulations and found that the Jeans equation could provide a reasonable approximation to the system dynamics if the amplitude of the radial velocity component is significantly smaller than the azimuthal one. It is possible to assume an axisymmetric steady disc to measure the RC globally using the Jeans equation. In Figure~\ref{fig:rc_compa} (see Appendix~\ref{AppendixA}), we also compared the RC with different limits for the vertical heights and they show good consistency even in the outer disc. The warp could affect spatial and velocity measurements, and especially the vertical velocities, which are expected to be its main kinematic signature. In order to measure the effect of the vertical velocity, we analysed the neglected cross-term $\langle V_RV_z\rangle$ in detail (see Sect.~\ref{sec:syst_ct}). We also note that \citet{wang2023} applied LIM to the full \gaia DR3 data set with different stellar populations, which helps to reduce the warp impact because we know that the warp is much stronger for young populations. 

\subsection{Have we reached the Keplerian decline of the Milky Way rotation curve?}

Our RC measurements based on \gaia DR3 (as well as that from \cite{ou2023}) demonstrate its significant decline. Additionally, the slope of the declining RC is steeper compared to the previous studies \citep{eilers2019,mroz2019} based on \gaia DR2. This also implies that the increase in cumulative mass at larger radii is minimal. We apply an approximate conversion from the measured circular velocity to the enclosed dynamical mass using the following relation:
\begin{equation}
    M_{\mathrm{dyn},i}=\frac{v_{\mathrm{circ},i}^2 R_i}{G}
    \label{eq:mass_grav}
,\end{equation}

\begin{figure}
 \centering
 \includegraphics[width=1\columnwidth]{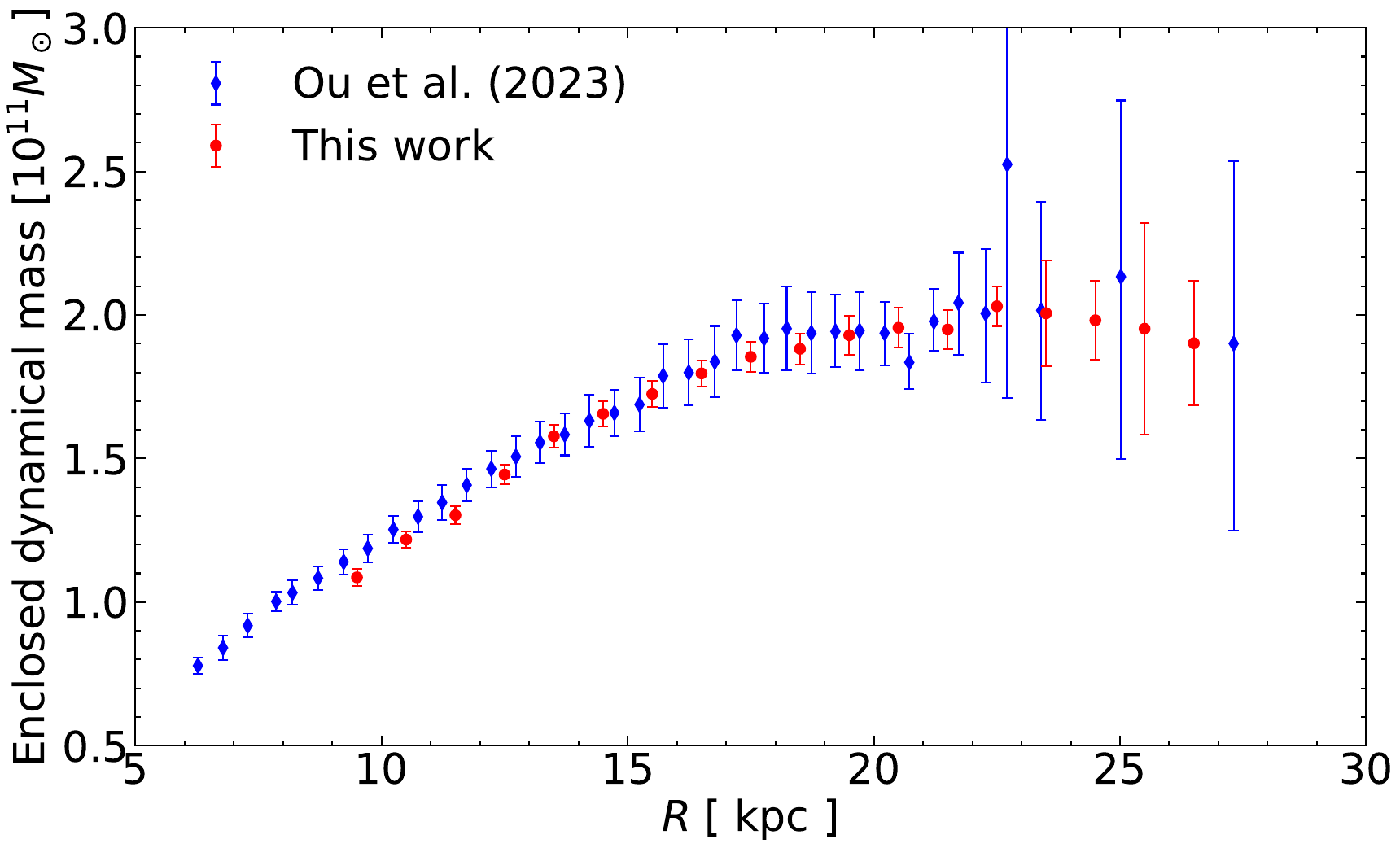}
    \caption{Enclosed dynamical mass of the MW as a function of radius.}
\label{fig:mass_R}
\end{figure}

where $G$ is the gravitational constant and $v_{\mathrm{circ},i}$ is the circular velocity at each radius $R_i$. The conversion is not accurate in a non-spherical potential but the difference is very small at large radii, where the spherical DM component dominates (see Figure~\ref{fig:rc_fit}).  By applying Eq.~\ref{eq:mass_grav}, we determined that beyond $R>19$ kpc, the enclosed mass barely varies (see Figure~\ref{fig:mass_R}), remaining within 1.9 to 2.0 $\times$ $10^{11}M_{\sun}$, which is remarkably close to our estimate made at a significantly larger radius (see Table~\ref{tab:mass}). The small decay of enclosed dynamical mass at large radii ($R>23$ kpc) in Figure~\ref{fig:mass_R} cannot be physical. However, given that the amplitude of the decay is much smaller than the error bars, it has no incidence on the validity of this work.


\begin{figure}
    \centering
    \includegraphics[width=\columnwidth]{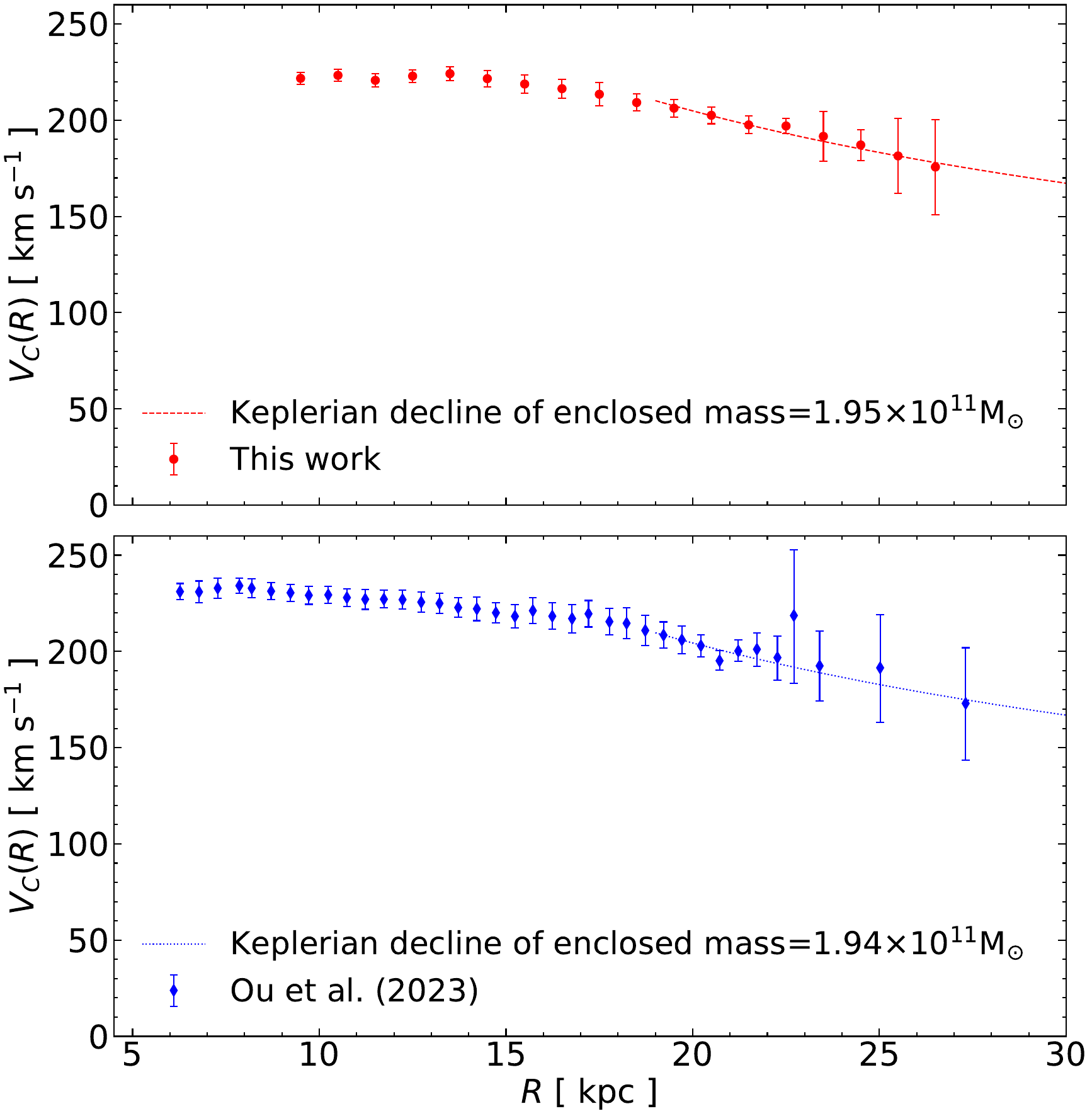}
    \caption{Rotation curve with the best fit of the Keplerian decline for our RC measurement (top panel) and that from \citet{ou2023} (bottom panel).}
    \label{fig:rc_kepl}
\end{figure}

Given that the RC drops faster at large radii and because of the agreement between enclosed and virial masses, we conducted a test to assess whether or not the MW RC has reached a Keplerian decline beyond $R>19$ kpc. In Figure \ref{fig:rc_kepl}, we compare the Keplerian decline using the same enclosed mass at each radius beyond $R>19$ kpc, which helps us to verify whether or not the MW RC at the outskirts can be fitted this way. Figure \ref{fig:rc_kepl} displays the best fit for a Keplerian decline to the RCs from our measurement (top panel) and that of \citet[bottom panel]{ou2023}, respectively. At large radii, this fits well with the two estimated RCs, suggesting that stellar rotation at the outskirts follows a Keplerian decline, which implies that the enclosed mass within a certain radius is sufficient to account for the observed velocities, without requiring an increase in mass at larger radii. Interestingly, we find a consistent enclosed mass for the Keplerian decline in both RC measurements, which amounts to approximately $1.95 \times 10^{11} M_{\sun}$. \\

\begin{figure*}
    \centering
\begin{minipage}{\columnwidth}
    \centering
    \includegraphics[width=\columnwidth]{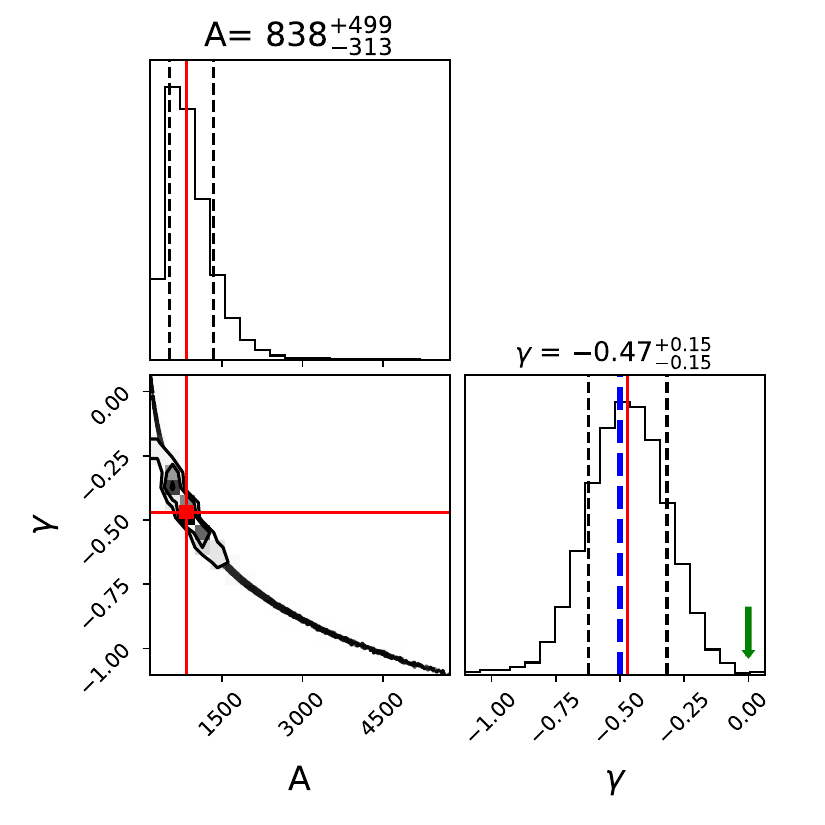}
\end{minipage}
\begin{minipage}{\columnwidth}
    \centering
    \includegraphics[width=\columnwidth]{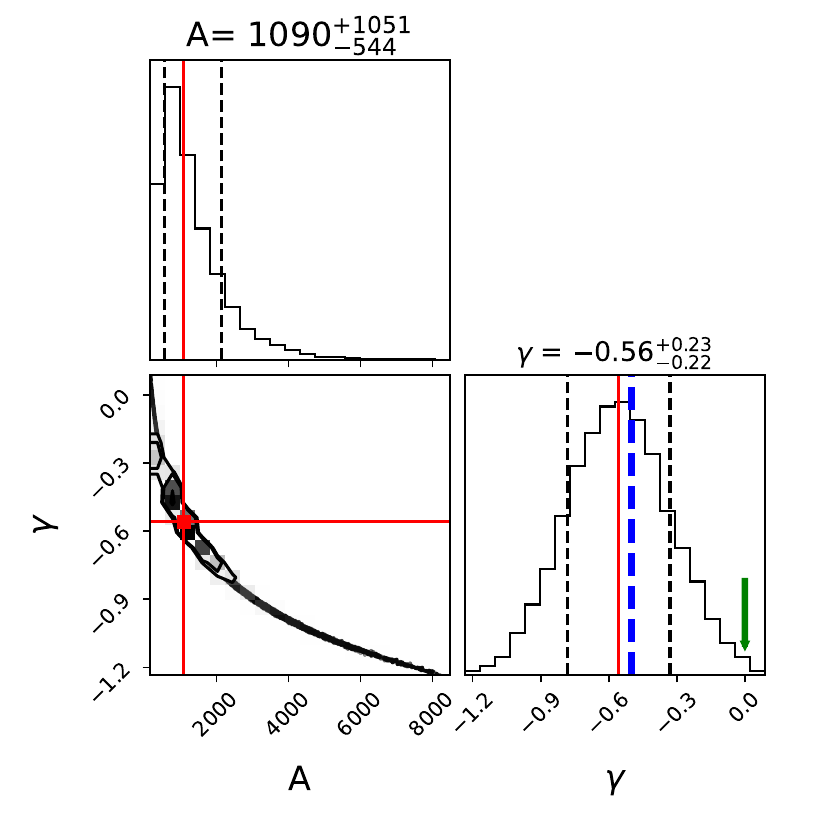}
\end{minipage}
    \caption{Posterior distributions for the parameters $A$ and $\gamma$. We show fits to the RC of this work (left panel) and to that of \citet{ou2023} (right panel) including systematics uncertainties. Histograms of parameters $A$ and $\gamma$ correspond to the maximum density distribution. Red solid lines represent the median of the posterior distributions. Black dashed lines correspond to $1\sigma$ uncertainty. Blue dashed lines correspond to the slope of the Keplerian decline, $n=-0.5$. Green arrows indicate the slope of a flat RC ($n=0$).}
    \label{fig:mcmc_kepl}
\end{figure*}

In order to further test for the presence of a Keplerian decline at large radii, we assumed a circular velocity profile following:
\begin{equation}
    V(R)=AR^\gamma
,\end{equation}
where A is an amplitude parameter, and $\gamma$ is the exponential slope of the outer RC. The posterior distributions are presented in Figure~\ref{fig:mcmc_kepl}. By sampling $\gamma$ using a flat prior between $-10$ and $5$, we found that the slope $\gamma$ beyond $R>19$ kpc is $-0.47^{+0.15}_{-0.15}$ and $-0.56^{+0.23}_{-0.22}$ for the RC of this work and that of \cite{ou2023}, respectively. This suggests that the MW RC is consistent with a Keplerian decline ($\gamma$=-0.5) at large radii, while a flat RC is rejected at a $3\sigma$ significance level (see green arrows in Figure~\ref{fig:mcmc_kepl}).

\subsection{Comparison of the MW dynamical mass with that based on different mass tracers}

The RC of the Milky Way is the most accurate tracer for estimating the enclosed dynamical mass within a range of approximately $30$ kpc, because rotating stars in the disc are likely at equilibrium with the total potential associated with the enclosed dynamical mass. We note that our dynamical mass estimate is primarily constrained by the RC within the range of $R=9-27$ kpc and is then extrapolated to the virial radius. The extrapolation is unlikely to bring an additional mass component, because the MW RC seems to follow a Keplerian decline at large radii. \\
Other estimates derived from stellar streams \citep{vasiliev2021,koposov2023}, globular clusters \citep{eadie2019, posti2019}, the Magellanic Cloud \citep{magnus2022}, or dwarf galaxies \citep{cautun2020,li2020, slizewski2022} provide virial masses ranging from 7 to 11$\times 10^{11}\,M_{\sun}$ at about 150 to 250 kpc. Our predicted virial mass from the MW RC is then considerably lower than these previous estimates.\\ 
However, the findings of recent studies of the orbital histories of halo inhabitants may challenge our previous understanding. \cite{hammer2023} found that most dwarf galaxies are relative newcomers. This is due to the expected linear relationship between infall times and the logarithm of the total orbital energy \citep{Rocha2012}. As dwarf-galaxy orbital energies are larger than those of former events such as the GSE and the Sgr infall, they likely entered the MW halo less than 3 Gyr ago. Moreover, \cite{li2021} found that the dwarf galaxies are highly concentrated near their pericentre. Consequently, the assumption of virial equilibrium for dwarf-galaxy orbital motions is likely invalid, and it is not surprising that one finds very large mass values when using these orbital motions as mass tracers for the MW. This is well illustrated by \citet{Boylan-Kolchin2013} who showed that assuming Leo I as a bound satellite would lead to a significant overestimate of the MW mass.\\
After using both globular clusters and the MW RC from \gaia DR2, \cite{wang2022} found that excluding Crater and Pyxis, which possess large amounts of orbital energy, leads to a decrease in the MW mass estimate from $5.73^{+0.76}_{-0.58}\times 10^{11} M_{\sun} $ to $5.36^{+0.81}_{-0.68}\times 10^{11} M_{\sun}$ when assuming an Einasto profile. \\
These recent findings highlight the fact that our understanding of the MW structure is changing. They also emphasise the need to verify whether or not various tracers have their orbital velocity in equilibrium with the MW potential in order to avoid systematic overestimates of the total mass.

\subsection{Is our Galaxy exceptional or is its rotation curve related to the \gaia methodology used to recover it?}

We investigated the literature on searching for declining spiral galaxy RCs to find out whether or not some of them may show a Keplerian decline. In reviewing galaxy discs, \citet{vanderKruit2011} mentioned that none spiral galaxies show a decline in their RCs, which can be associated with a cut-off in the mass distribution, so that in no case has the RC been traced to the limit of the dark matter distribution. In their analysis of well-studied spirals from the THINGS project, \citet{deBlok2008} found that a declining RC such as that of M81 is likely caused by galaxy interactions. \\

\citet[see also \citealt{Zobnina2020}]{Noordermeer2007} specifically studied spirals with a declining RC and found no RCs with a fully Keplerian decline in the outer regions, indicating that we have not yet reached the point where the mass density becomes negligible, except perhaps for UGC 4458 (NGC 2599). However, the latter shows a bright UV nucleus that may be active, and its disc morphology also appears quite different from that of the MW.  \citet{Dicaire2008} identified NGC 7793 as a possible candidate for exhibiting a Keplerian decline; its DM content is surprisingly low for a dwarf galaxy, and is smaller than that of stars at all radii. \\

The MW may indeed be relatively exceptional if it is the only isolated grand-design spiral showing a Keplerian decline in its RC, which is consistent with a possible cut-off of its mass distribution. One might wonder whether or not this declining RC is related to its four-arm structure discovered by \citet{Georgelin1976}. Four-armed spirals are rare in nearby Milky Way-type galaxies, although galaxies with two arms frequently exhibit bifurcations resulting in several arm segments, which could also be the case for the Milky Way. \\

Alternatively, our Galaxy may be exceptional due to its relatively quiet past history \citep{Hammer2007,Belokurov2018,Haywood2018}, having experienced no major merger for $\sim$ 9 Gyr. We must also consider the possibility that the methodology used by \gaia to recover its full 6D space-velocity parameters for very large numbers of stars may be a contributing factor to the exceptional nature of the RC we obtain based on these data; it certainly contrasts with the less constrained RC of external galaxies.

\section{Conclusions}

\gaia DR3 has led to significant progress in our capacity to estimate the MW RC when compared to \gaia DR2. Three different studies have derived the RC of the MW: One used the whole and very large \gaia DR3 catalogue with distances estimated from parallaxes, and then averaged into specific 6D cells using LIM \citep{wang2023}. The second \citep{ou2023} is based on a smaller number of RGB stars (33 335), for which distances are estimated from spectrophotometry. The third \citep{zhou2023} is based on 58 000 bright RGB stars, and in the present study we re-evaluated their distance estimates, rendering their RC consistent with that of the two other studies.\\

Here, we carried out a full analysis of the systematic uncertainties that can affect the \citet{wang2023} study. We then compared the different RCs from \gaia DR3, and find that they have consistent rotational velocity values from 9 to 27 kpc. This indicates a robust and significant decline of the MW RC, with a slope of $\beta=-(2.18\pm 0.23)\rm\,km\,s^{-1}\,kpc^{-1}$, which is based on combining the very similar slopes from this study and that from \citet{ou2023}. The decrease in velocity between 19.5 and 26.5 kpc is approximately 30 km\,s$^{-1}$. \\

The estimated MW dynamical mass is consistent with  $1.99^{+0.09}_{-0.06}\times 10^{11}\ M_{\sun}$ at $121.03^{+1.80}_{-1.23}$ kpc (from the RC of this study) and with $2.13^{+0.17}_{-0.12}\times 10^{11}\ M_{\sun}$\ at $123.80^{+3.21}_{-2.37}$ kpc (from the RC of \citealt{ou2023}), respectively\footnote{We note that the two estimates are very similar and the average value would be $2.06^{+0.24}_{-0.13}\times 10^{11}\ M_{\sun}$}. The relatively small size of the error bars is perhaps a result of the assumption that the two RC points and their associated error bars were independently determined. \\
 
Consequently, the ratio of DM to baryonic mass is only a factor of about 3, instead of a factor of approximately 6 from $\Lambda$CDM \citep{Planck2020}, which suggests that baryons are not missing in our Galaxy. A small dynamical mass for the MW may also impact mass estimations for the  LMC, which is mostly constrained by its ratio to that of the MW; for example, when studying the induced sloshing of the Galactic halo \citep{Erkal2021,Conroy2021}. If the total MW mass is as small as $2.06 \times 10^{11}\ M_{\sun}$, the LMC total mass would be from 2 to $3\times 10^{10} M_{\sun}$. Interestingly, the latter value is consistent with the modelling of the Magellanic Stream through ram pressure as shown by \citet{Hammer2015} and \citet[see also LMC mass predictions from \citealt{Wang2022b}]{Wang2019}, which would resolve the numerous difficulties in reproducing it with tidal tail models \citep{Besla2012,Lucchini2020,Lucchini2021}.\\

We conclude that the MW RC cannot be consistent with a flat RC at a significance of $3\sigma$, and that our findings suggest a Keplerian decline occurring at radii of greater than 19 kpc. The Keplerian decline indicates the point where the mass density becomes negligible \citep{Zobnina2020}. Some spiral galaxies present a declining RC, but at large radii they appear to flatten out, meaning that their RC has not been traced to the outermost extent of the dark matter distribution \citep{deBlok2008,vanderKruit2011}. This contrasts with the MW, whose accretion history \citep[and references therein]{hammer2023} shows no major merger for 8 to 10 Gyr, while half of the spiral galaxies underwent their last major merger more recently \citep{Hammer2005,Hammer2009, Puech2012}. It would be interesting to study the impact of relatively recent assembly events on the RC at the outskirts of spiral galaxies. \\

In many respects, a Keplerian decline for the MW RC may appear quite exceptional. This could be due to the extraordinarily quiet history of our Galaxy, or to the very different methodology used by \gaia to calculate its kinematics compared to that used to study external galaxies.

\begin{acknowledgements}
      We thank the anonymous referee for their comments and suggestions. Y.-J.J. acknowledges financial support from the China Scholarship Council (CSC) No.202108070090. J.-L.W. is also grateful to be supported by the CSC No.202210740004. L.C. acknowledges the Chilean Agencia Nacional de Investigacion y Desarrollo through the grant Fondecyt Regular 1210992. We are grateful to Chrobáková Žofia, López-Corredoira Martín and Francesco Sylos Labini for their invaluable assistance in producing the datasets and tests of \cite{wang2023}. We thank Yang Huang for sharing their data from \cite{zhou2023}. We are grateful for the support of the International Research Program Tianguan, which is an agreement between the CNRS in France, NAOC, IHEP, and the Yunnan Univ. in China. The data underlying this article will be shared on request to the corresponding author.
\end{acknowledgements}




\bibliographystyle{aa}
\bibliography{main} 




\begin{appendix}

\section{The impact of the z selection on the RC between 9 and 13 kpc}
\label{AppendixA}
\begin{figure}
\centering
    \includegraphics[width=\columnwidth]{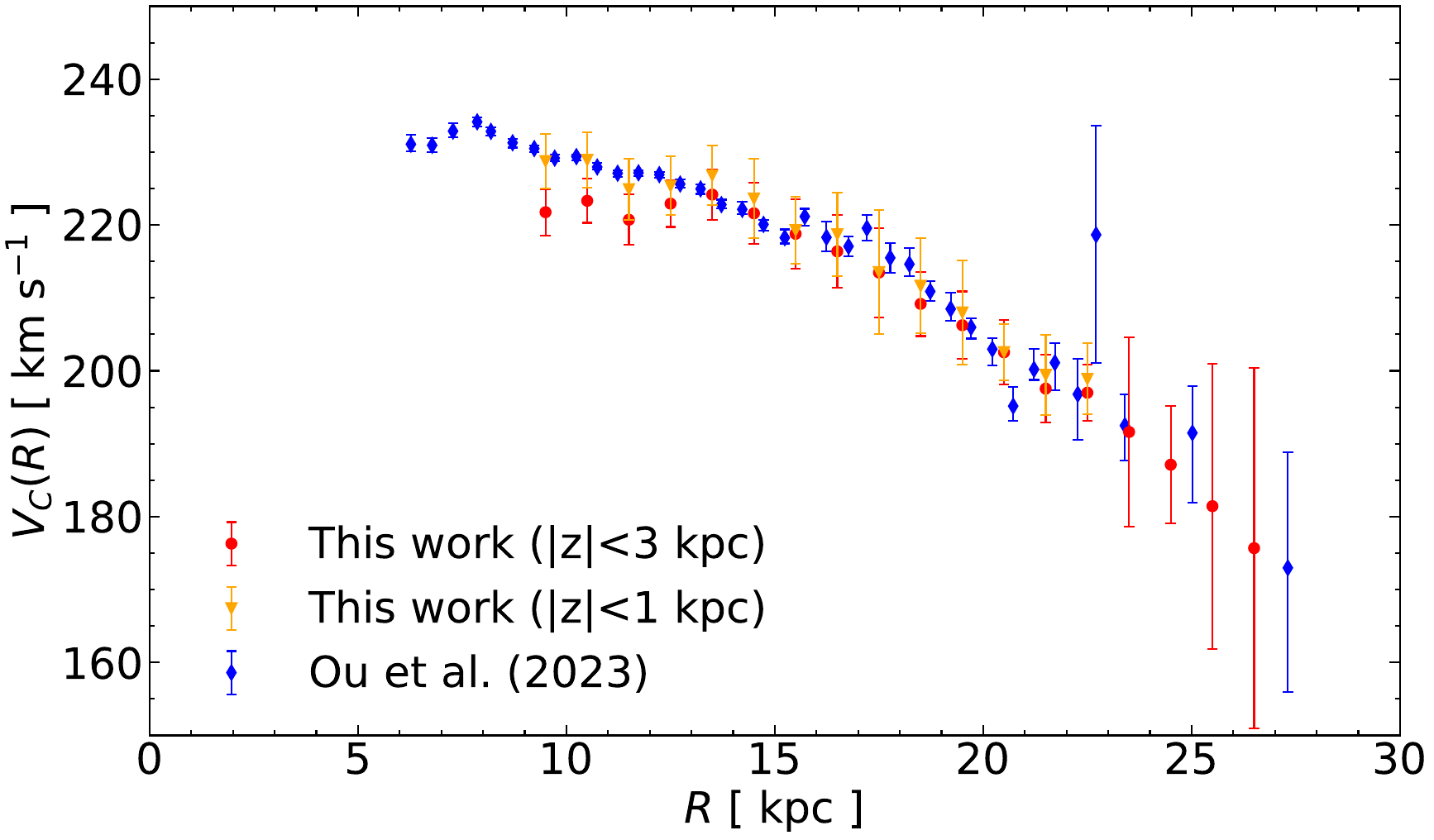}
    \caption{Comparison of the RC using different selections in the vertical direction.}
    \label{fig:rc_z1}
\end{figure}

Figure~\ref{fig:rc_z1} indicates that by limiting the \cite{wang2023} data to $|z| < $ 1 kpc, their RC from 9 to 13 kpc becomes consistent with that of \citet{ou2023}. 

\newpage

\section{Comparison of distance estimates}
\label{AppendixB}

\begin{figure*}
    \centering
    \includegraphics[width=\textwidth]{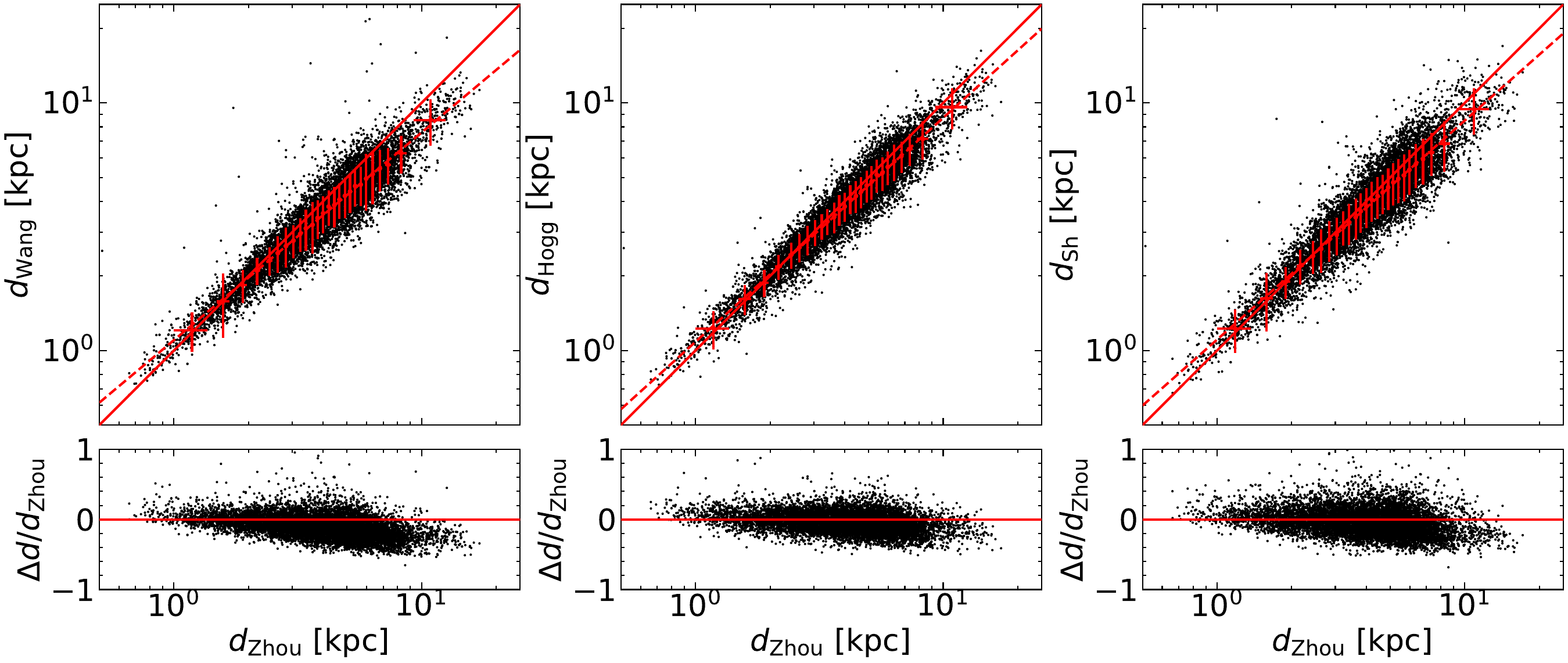}
    \caption{Comparison of distance estimates made by \citet{zhou2023} to those made by \citet{wang2016}, \citet{Hogg2019}, and StarHorse \citep{queiroz2023} in the top panels, respectively. The solid line presents a one-to-one correspondence. The dashed line presents the best linear fit on the logarithmic scale. Red points and error bars indicate the median and standard deviation per 400 data points. The ratios $\Delta d/d_\mathrm{zhou}$ are shown in the bottom panels.}
    \label{fig:dist_compa}
    \end{figure*}

\begin{figure}
    \centering
    \includegraphics[width=\columnwidth]{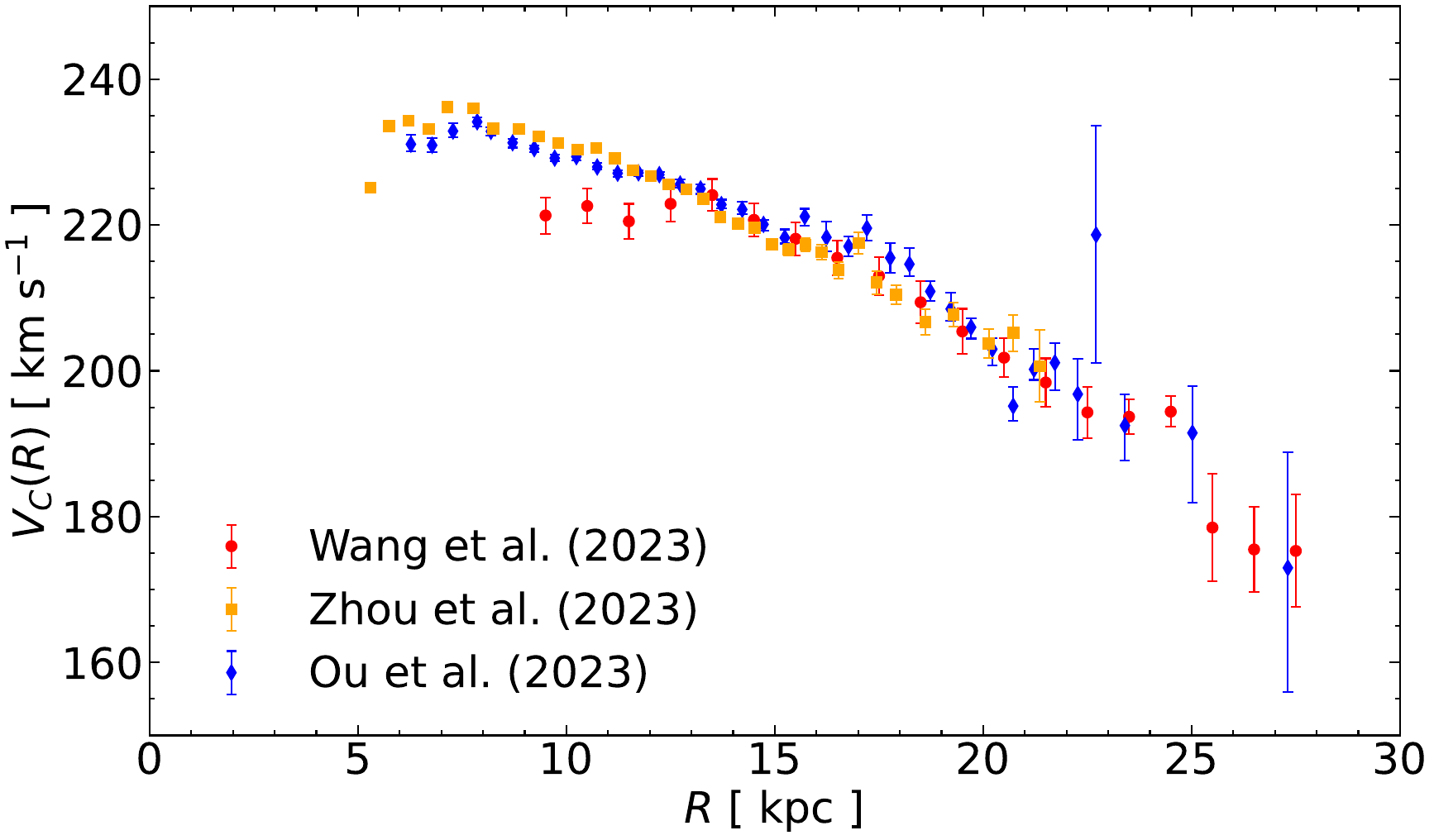}
    \caption{Comparison of RCs after a rough correction of the distance applied to the points from \citet[see text]{zhou2023}.}
    \label{fig:corr_Zhou}
\end{figure}

Figure~\ref{fig:dist_compa} compares the distance estimates by \citet{zhou2023} with those of three different methods on the APOGEE data. One is a modified version of \cite{wang2016}, in which \gaia EDR3 parallax has been incorporated as a prior to constrain stellar distances as described in \cite{wang2023dis}. Other distance estimates are from \citet[which were used by \citealt{eilers2019}]{Hogg2019}, and from the results of StarHorse \citep{queiroz2023}. This comparison shows that \citet{zhou2023} always overestimate distances for R $>$ 10 kpc stars. To evaluate the consequences of such a bias, Figure~\ref{fig:corr_Zhou} provides a very rough analysis by correcting each  \citet{zhou2023} RC point accordingly, which suffices to reconcile their RC with those from this study and from \citet{ou2023}. However, a better analysis, where the distances are corrected for individual stars, is still needed.

\end{appendix}


\end{document}